\documentclass[reprint,onecolumn,floatfix,nofootinbib,aps,prd,superscriptaddress,amsmath,amssymb,11pt]{revtex4-2}

\usepackage{graphicx}
\usepackage{dcolumn}
\usepackage{bm}
\usepackage{braket}
\usepackage{mathrsfs}
\usepackage{color}
\usepackage[colorlinks=true,linkcolor=red,citecolor=blue]{hyperref}
\newcommand{\revisedtext}[1]{{\bfseries\boldmath #1}}

\begin{document}
\title{\texorpdfstring{Horizon-Evanescent Scalar Clouds from Coupled Rotation and Magnetic Fields\\ around Black Holes}{Horizon-Evanescent Scalar Clouds from Coupled Rotation and Magnetic Fields around Black Holes}}

\author{Hengyu Xu}
\email{xuhengyu0501@outlook.com}

\author{Haowei Chen}
\email{chenhaowei@zjut.edu.cn}

\author{Shao-Jun Zhang}
\email{sjzhang@zjut.edu.cn (corresponding author)}

\affiliation{Institute for Theoretical Physics and Cosmology$,$ Zhejiang University of Technology$,$ Hangzhou 310032$,$ China}
\affiliation{School of Physics and Optical Engineering$,$ Zhejiang University of Technology$,$ Hangzhou 310032$,$ China}
\date{\today}

\begin{abstract}
We show that black-hole rotation and an external magnetic field can jointly generate a distinct class of scalar clouds. Using the Kerr-Bertotti-Robinson geometry as a separable laboratory for magnetized rotating black holes, we study a charged massive scalar field and map the radial Klein-Gordon equation into a one-dimensional Schr\"{o}dinger-like form. The magnetic coupling shifts the near-horizon dispersion relation and realizes a positive horizon gap: a sufficient near-horizon criterion under which the horizon wavenumber becomes purely imaginary in a finite frequency band below the usual kinematic synchronization frequency. In this band the physical horizon boundary condition is no longer a propagating ingoing wave, but a regular exponentially decaying state. This rotation--magnetic-field mechanism quenches the superradiant flux and supports horizon-decaying scalar clouds (Type-II), distinct from the usual synchronized propagating clouds (Type-I). Matched asymptotic expansions and numerical shooting solutions are used to exhibit both branches and their spatial profiles. Thus the Kerr-Bertotti-Robinson solution is not an isolated curiosity, but an explicit realization of a broader positive-gap criterion for horizon-evanescent stationary configurations that are absent in isolated Kerr systems.
\end{abstract}

\maketitle

\section{Introduction}
\label{sec:intro}

The extraction of rotational energy from a black hole is one of the most fascinating predictions of general relativity, initially conceptualized by Penrose \cite{Penrose:1969pc, christodoulou1970reversible, ruffini1971introducing}. In the realm of wave dynamics, this energy extraction manifests as superradiance. When a bosonic wave impinges on a rotating black hole, it can be amplified if its frequency $\omega$ satisfies the superradiant condition $0 < \omega < m\Omega_H$, where $m\in\mathbb{Z}$ is the azimuthal quantum number and $\Omega_H$ is the angular velocity of the black hole horizon \cite{Zeldovich:1971ffh, Zeldovich:1972zqp, misner1972interpretation, Starobinskii:1973vzb, bekenstein1973extraction, teukolsky1973perturbations}. 

If the bosonic field possesses a non-zero rest mass $\mu_s$, the mass term acts as a natural effective mirror at spatial infinity, trapping the amplified low-frequency modes in the vicinity of the black hole \cite{press1972floating}. This geometric confinement leads to a continuous, exponentially growing extraction of energy, triggering the well-known superradiant instability, often referred to as the ``black hole bomb'' mechanism \cite{damour1976black, zouros1979instabilities, detweiler1980klein, furuhashi2004instability}. Over the decades, the massive scalar field instability has been thoroughly investigated through analytical matching techniques and precise numerical methods across various regimes of the parameter space \cite{cardoso2004black, dolan2007instability, dolan2013superradiant, witek2012superradiant}; see Ref.~\cite{Brito:2015oca} for a comprehensive review.

At the exact threshold of this superradiant instability ($\omega = m\Omega_H$ in Kerr, or its charged generalization in Kerr-Newman), the net energy flux across the event horizon strictly vanishes. In this highly tuned kinematic regime, the massive field does not grow or decay, but instead forms a completely stationary, non-dissipative bound state. These long-lived configurations, famously known as ``scalar clouds,'' have been extensively studied in standard Kerr and Kerr-Newman spacetimes \cite{hod2012stationary, hod2014kerr, herdeiro2014kerr, herdeiro2015construction}. They represent the linear onset of black hole scalar hair, challenging the traditional no-hair paradigm and providing useful insight into the non-linear dynamics of fields in curved spacetimes \cite{benone2014kerr, Huang:2017whw, Herdeiro:2015waa}. Analogous bound-state phenomena have also been explored for vector and tensor fields \cite{pani2012black, PhysRevLett.119.041101, herdeiro2016kerr}, as well as in acoustic black hole analogues \cite{hod2021stationary, ciszak2021acoustic}. In all these familiar cases, however, the zero-flux cloud is still obtained as the synchronized limit of a propagating horizon wave. The possibility that a stationary cloud may instead be supported by an evanescent, horizon-decaying boundary condition has not been systematically explored.

Astrophysical and theoretical black holes are often considered in environments containing electromagnetic fields, for example those generated by accretion disks and diffuse plasmas \cite{blandford1977electromagnetic, wald1974black}. The presence of an external magnetic field significantly alters both the spacetime geometry and the effective interactions of any charged fields propagating within it. Traditionally, the exact solutions describing rotating black holes immersed in strong magnetic universes have been constructed via the Ernst formulation \cite{ernst1976black,ernstwild1976kerr}, which relies on the Melvin magnetic universe \cite{melvin1964pure}. The thermodynamic and geometric properties of such magnetized rotating black holes can differ substantially from asymptotic flatness, introducing complex ergoregion structures and altered horizon topologies \cite{gibbons2013ergoregions,booth2015insights}. Naturally, this exact framework has been the primary arena for investigating how magnetic fields affect superradiant instabilities. Studies of wave dynamics in Ernst-type backgrounds have revealed intricate couplings between the field's charge and the external magnetic potential, demonstrating that magnetic fields can dynamically enhance or suppress superradiance \cite{Konoplya:2008hj,brito2014superradiant}, and explicitly supporting the existence of stationary scalar clouds in Melvin-Kerr spacetimes \cite{santos2021black}.

Despite these analytical successes, investigating wave dynamics in the Ernst formulation poses a significant mathematical challenge: the standard magnetized Kerr solution generally loses the algebraic Type D classification of the original isolated Kerr metric. This loss of algebraic symmetry severely complicates the exact separability of the Klein-Gordon and Teukolsky equations, often forcing analyses of scalar clouds to rely on stringent weak-field limits, extreme spin assumptions, or specialized numerical approximations. To address this mathematical bottleneck and study magnetized scalar clouds in a more geometrically tractable framework, one can turn to alternative exact solutions of the Einstein-Maxwell equations. 

Recently, Podolsk\'{y} and Ovcharenko formulated an exact spacetime---the Kerr-Bertotti-Robinson (Kerr-BR) black hole \cite{podolsky2025kerr, PhysRev.116.1331, robinson1959solution}, whose rigorous thermodynamic properties, conserved charges, and first law of black-hole thermodynamics have also been recently established \cite{Hu:2026slp}. In the present work, we use this geometry primarily as a controlled theoretical laboratory for isolating how rotation and a magnetic field act together on scalar-field boundary conditions. To build physical intuition, it is useful to distinguish this exact geometry from the well-known Wald solution \cite{wald1974black}. While the Wald solution merely superimposes a test magnetic field onto a standard asymptotically flat Kerr vacuum, the Kerr-BR metric fully incorporates the non-linear gravitational backreaction of the electromagnetic field. Consequently, its asymptotic structure does not reduce to flat space, but naturally approaches the Bertotti-Robinson universe---a geometry intrinsically supported by a uniform magnetic field. Furthermore, unlike the Ernst-magnetized models, this spacetime successfully retains the algebraic Type D classification while describing a Kerr black hole immersed in an asymptotically uniform magnetic field with non-aligned principal null directions. This algebraic structure makes the weak-field separable sector especially useful for identifying mechanisms that would otherwise be obscured by non-separability.

In this paper, we systematically investigate the dynamics of a massive charged scalar field in the weak-field separable sector of the Kerr-BR spacetime. The central result is a rotation--magnetic-field mechanism for producing a second, horizon-decaying class of scalar cloud. The rotation supplies the usual superradiant channel through the azimuthal coupling to the horizon angular velocity, while the magnetic field dresses both the asymptotic confinement and the near-horizon dispersion relation of the charged scalar. Their combined effect creates a finite frequency interval in which the mode remains below the kinematic synchronization frequency, but the effective radial wavenumber at the horizon becomes imaginary. The horizon boundary condition therefore changes from a propagating ingoing wave to a regular evanescent state. Through matched asymptotic expansions using confluent hypergeometric functions \cite{abramowitz1964handbook, bender1999advanced}, we show that this transition quenches the superradiant flux and removes the dissipative imaginary part of the bound-state frequency.

Consequently, the usual synchronized scalar clouds appear here as Type-I states located at the lower boundary of the propagating branch, whereas the horizon-decaying branch consists of Type-II scalar clouds that exponentially decay toward the event horizon. Supported by numerical shooting integrations \cite{press2007numerical}, we show that these Type-II clouds have a near-horizon structure that is qualitatively different from the plateau behavior of the synchronized clouds.

The main message of this work can be summarized in three points. First, the zero-flux condition for a scalar cloud need not arise only from synchronization of a propagating horizon wave. Second, in a magnetized rotating geometry the combined rotational and electromagnetic couplings can open a positive near-horizon gap, turning the horizon channel into an evanescent one. Third, once this evanescent branch is combined with confinement at infinity, it supports a stationary scalar cloud whose horizon behavior is qualitatively different from the familiar Kerr and Kerr-Newman clouds. In this sense, the Kerr-BR spacetime is not the end point of the analysis, but a clean realization of a more general positive-gap criterion: whenever the separated near-horizon radial equation contains a positive gap term, an evanescent scalar-cloud branch becomes possible.

The paper is organized to emphasize this mechanism. In Sec.~\ref{sec:spacetime_kg}, we introduce the Kerr-BR spacetime as a tractable magnetized rotating background and derive the scalar-field equation. In Sec.~\ref{sec:effective_potential}, we map the radial equation into a Schr\"{o}dinger-like form, derive the horizon gap, and state a general criterion for horizon-evanescent clouds. Sec.~\ref{sec:analytic_quenching} presents the analytic bound-state argument for the rotation--magnetic quenching mechanism. In Sec.~\ref{sec:numerical}, we present numerical shooting results that distinguish the propagating Type-I reference clouds from the horizon-decaying Type-II clouds. Finally, we conclude in Sec.~\ref{sec:conclusions}.

\section{A Tractable Magnetized Rotating Background}
\label{sec:spacetime_kg}

To investigate the rotation--magnetic-field mechanism in a setting where the scalar equation remains analytically tractable, we consider a Kerr black hole immersed in an external uniform magnetic field, described by the Kerr-Bertotti-Robinson or Kerr-BR spacetime \cite{podolsky2025kerr}. The metric has a compact explicit form
\begin{equation}
ds^2 = \frac{1}{\Omega^2} \left[ - \frac{Q}{\rho^2} \left( dt - a \sin^2 \theta \, d\varphi \right)^2 + \frac{\rho^2}{Q} \, dr^2 + \frac{\rho^2}{P} \, d\theta^2 + \frac{P}{\rho^2} \sin^2 \theta \, (a \, dt - (r^2 + a^2) \, d\varphi)^2 \right],
\label{eq:metric}
\end{equation}
where the metric functions are given by
\begin{equation}
\begin{aligned}
\rho^2 &= r^2 + a^2 \cos^2 \theta, \\
P &= 1 + B^2 \left( M^2 \frac{I_2}{I_1^2} - a^2 \right) \cos^2 \theta, \\
Q &= \left( 1 + B^2 r^2 \right) \Delta, \\
\Omega^2 &= \left( 1 + B^2 r^2 \right) - B^2 \Delta \cos^2 \theta,
\end{aligned}
\end{equation}
with the horizon function $\Delta = \left( 1 - B^2 M^2 \frac{I_2}{I_1^2} \right) r^2 - 2M \frac{I_2}{I_1} \, r + a^2$. Here, $M$ and $a$ represent the mass and specific angular momentum of the black hole, respectively, and $B$ denotes the magnetic field strength. The constants are $I_1 = 1 - \frac{1}{2} B^2 a^2$ and $I_2 = 1 - B^2 a^2$.

The electromagnetic field is given by the complex 1-form potential (its real counterpart is $A^{\text{real}} \equiv 2 \, \text{Re} \, A$):
\begin{equation}
A = \frac{e^{i \xi}}{2B} \left[ \Omega_{,r} \frac{a \, dt - (r^2 + a^2) \, d\varphi}{r + i a \cos \theta} + \frac{i \Omega_{,\theta}}{\sin \theta} \frac{dt - a \sin^2 \theta \, d\varphi}{r + i a \cos \theta} + (\Omega - 1) \, d\varphi \right],
\end{equation}
which has a smooth $B\to0$ limit and vanishes when the magnetic field is removed.

The propagation of a massive, charged test scalar field $\Phi$ with mass $\mu_s$ and charge $q$ is governed by the covariant Klein-Gordon (KG) equation:
\begin{equation}
\left[ \left( \nabla_\mu - iqA_\mu \right) \left( \nabla^\mu - iqA^\mu \right) - \mu_s^2 \right] \Phi = 0. \label{eq:kge}
\end{equation}
When the magnetic field is weak, we use the dimensionless convention $M=1$ for the displayed parameter ranges, so that $B \ll 1$ is equivalent to the unscaled condition $|B|M \ll 1$. We explicitly take the electromagnetic phase parameter to be $\xi = 0$, which represents the pure magnetic field. We work in the weak-field separable sector of the geometry, retaining the leading magnetic couplings that enter the charged scalar dynamics while keeping the radial and angular equations decoupled.

At generic $B$, Eq.~\eqref{eq:kge} is not separable for a massive charged
scalar.  The Kerr--BR metric is conformal to a Carter metric, with conformal
factor $\Omega(r,\theta)$ defined in Eq.~\eqref{eq:metric}.  The massless
conformally coupled equation consequently admits generalized conformal
separation \cite{Gray2026}; because the Kerr--BR Ricci scalar vanishes, this
includes the uncharged massless minimally coupled equation.  A nonzero mass
introduces a fixed scale; under the conformal map, the mass term becomes
$\mu_s^2\Omega^{-2}(r,\theta)$ and mixes $r$ and $\theta$.  Hence $q=0$ is
insufficient when $\mu_s\ne0$.
For $q\ne0$, the non-aligned exact Maxwell field supplies additional mixing.
The weak-field truncation discards the higher-order mixing, so the separation
used below is approximate.

This controlled limit is sufficient for isolating the rotation--magnetic mechanism responsible for the horizon boundary condition. Under this condition, substituting the separable ansatz $\Phi(t, r, \theta, \phi) = \sum_{l,m} e^{i(m\phi - \omega t)} R(r) \, S(\theta)$ into the field equation allows us to derive the radial and angular equations. The complete forms of these equations, including the magnetic-field and scalar-charge couplings kept in this approximation, are detailed in Appendix \ref{app:full_equations}. The astrophysical relevance and limitations of this approximation are discussed in Sec.~\ref{sec:conclusions}.

\section{Near-Horizon Mechanism and General Criterion}
\label{sec:effective_potential}

\subsection{Schr\"{o}dinger-like Radial Equation in Tortoise Coordinates}

To analyze the physical boundary conditions, we map the radial equation into a one-dimensional Schr\"{o}dinger-like form. We introduce the tortoise coordinate $y$, mapping the exterior domain $r \in (r_+, \infty)$ to $y \in (-\infty, +\infty)$, defined by $dy/dr = (r^2 + a^2)/\Delta_0(r)$, where $\Delta_0(r) = r^2 - 2Mr + a^2$. Here, $r_\pm = M \pm \sqrt{M^2 - a^2}$ denote the outer and inner horizons, respectively.

By redefining the radial wave function as $Y = \sqrt{r^2 + a^2} \, R$, the radial equation is cast into:
\begin{equation}
\frac{d^2 Y(y)}{dy^2} - V_{\text{eff}}(y; \omega, \mu_s, B,q,a) \, Y(y) = 0. \label{eq:yreq}
\end{equation}
The properties of the bound states are completely dictated by the asymptotic behavior of $V_{\text{eff}}$. \revisedtext{The explicit expression for $V_{\mathrm{eff}}(r)$ within the retained weak-field approximation is given in Eq.~\eqref{eq:veff_explicit}.}

\subsection{Asymptotic Behavior and Physical Boundary Conditions}

\textbf{Spatial Infinity} ($y \to +\infty$, $r \to \infty$):
Far from the black hole, the effective potential approaches $k_f^2 \equiv B m q + \mu_s^2 - 2 a B q \omega - \omega^2$. In this asymptotic limit, the combination $\mu_{\text{eff}}^2 \equiv \mu_s^2 + B m q - 2 a B q \omega$ perfectly plays the role of an effective scalar mass squared, which is dynamically dressed by the external magnetic field. To investigate localized scalar clouds, we demand $k_f^2 = \mu_{\text{eff}}^2 - \omega^2 > 0$, imposing the bound-state requirement: $Y(y \to +\infty) \sim e^{-k_f y}$. 

To satisfy $k_f^2 > 0$, the bare scalar mass $\mu_s$ must be sufficiently large to counteract any repulsive effects from the magnetic coupling. While the complete piecewise constraints across all signs of $B$, $q$, and $m$ are collected in Appendix \ref{app:boundary_conditions}, the underlying physics becomes highly intuitive for the most relevant superradiant regime ($m > 0$, $a > 0$). As summarized in Table \ref{tab:mass_constraints}, when the magnetic field and scalar charge are aligned ($B q > 0$), the magnetic dressing enhances confinement, allowing even massless fields ($\mu_s \geq 0$) to form bound states. Conversely, an anti-aligned configuration ($B q < 0$) effectively reduces the scalar mass, thus necessitating a strict minimum bare mass cut-off to maintain wave confinement at spatial infinity.

\begin{table}[htbp]
\centering
\caption{Bound-state parameter constraints for the physically relevant superradiant regime ($m > 0, a > 0$). The field-charge alignment ($Bq$) dictates the minimum bare scalar mass ($\mu_s$) required for wave confinement. The exhaustive piecewise conditions for all $m, B, q$ limits are preserved in Appendix \ref{app:boundary_conditions}.}
\label{tab:mass_constraints}
\renewcommand{\arraystretch}{1.5}
\begin{tabular}{l @{\hspace{0.8cm}} c @{\hspace{0.8cm}} l}
\hline\hline
\textbf{Field-Charge Alignment} & \textbf{Azimuthal Mode} ($m>0$) & \textbf{Minimum Bare Mass} ($\mu_s$) \\
\hline
$B q > 0$ (Aligned) & All $m > 0$ & $\mu_s \geq 0$ \\
$B q = 0$ (Neutral/No Field) & All $m > 0$ & $\mu_s > 0$ \\
$B q < 0$ (Anti-aligned) & $0 < m \leq -a^2 B q$ & $\mu_s \geq 0$ \\
$B q < 0$ (Anti-aligned) & $m > -a^2 B q$ & $\mu_s \geq \sqrt{-a^2 B^2 q^2 - B m q}$ \\
\hline\hline
\end{tabular}
\end{table}

\textbf{Event Horizon} ($y \to -\infty$, $r \to r_+$):
Near the horizon, the effective potential asymptotes to $-k_h^2 \equiv - [ (\omega - \omega_c)^2 - (\omega_c^2 - D) ]$, where $\omega_c = m\Omega_H + q\Phi_H = ma/(2Mr_+) - aBqr_+/(4M)$ is the modified superradiant threshold frequency evaluated at the event horizon, with $\Omega_H$ and $\Phi_H$ being the angular velocity and electromagnetic potential at the horizon, respectively. The zeroes of the horizon wavenumber, $\omega_1 = \omega_c - \sqrt{\omega_c^2 - D}$ and $\omega_2 = \omega_c + \sqrt{\omega_c^2 - D}$, serve as critical phase-transition boundaries. 
The parameter $D$ captures the non-linear coupling:
\begin{equation}
D = -\frac{a^4 B m q + 2 m^2 M \left(\sqrt{M^2-a^2}-M\right) + a^2 m^2}{4 a^2 M^2}.
\end{equation}

To demonstrate the existence of this quenching band, we evaluate the discriminant $\omega_c^2 - D$. Substituting the explicit expressions for $\omega_c$ and $D$, and using $r_+=M+\sqrt{M^2-a^2}$, gives
\begin{equation}
\omega_c^2 - D = \frac{a^2 B^2 q^2 r_+^2}{16 M^2} > 0.
\label{eq:discriminant}
\end{equation}
For nonzero $a$, $B$, and $q$, this discriminant is positive. The critical phase-transition frequencies $\omega_1$ and $\omega_2$ are therefore real, and the quenching band $[\omega_1, \omega_2]$ exists in the physical parameter space considered here.

\subsection{Physical Horizon Boundary Condition}

The physical meaning of this band is fixed by the boundary condition at the future event horizon. A local observer just outside the horizon is dragged with angular velocity $\Omega_H$. Including the electromagnetic coupling, the near-horizon phase of a charged scalar mode is
\begin{equation}
\psi \sim e^{-i(\omega-m\Omega_H-q\Phi_H)t} e^{\pm i k_h y}
       = e^{-i(\omega-\omega_c)t} e^{\pm i k_h y}.
\label{eq:local_horizon_wave}
\end{equation}
For $k_h^2>0$, the horizon channel is propagating. The sign of the spatial phase is then chosen so that the wave is ingoing in the locally nonrotating frame \cite{bardeen1972rotating}. Thus the usual superradiant regime requires both the kinematic condition $0<\omega<\omega_c$ and a real horizon wavenumber. In the present system these two requirements do not define the same interval: the propagating superradiant branch ends at $\omega_1$, not at $\omega_c$.

For $\omega_1<\omega<\omega_2$, however, $k_h^2<0$ and the horizon channel is no longer propagating. Writing $k_h=-i\kappa_h$ with $\kappa_h>0$, the regular solution is
\begin{equation}
Y\sim e^{\kappa_h y},\qquad y\to-\infty ,
\label{eq:evanescent_bc}
\end{equation}
which decays toward the event horizon. Its conserved radial flux,
\begin{equation}
\mathcal{F}_H \propto \frac{1}{2i}\left(Y^*\frac{dY}{dy}-Y\frac{dY^*}{dy}\right),
\label{eq:horizon_flux}
\end{equation}
vanishes identically for the real exponential boundary condition. The quenching of superradiance is therefore not merely the usual synchronization condition; it is a change in the horizon channel itself, from a propagating wave to an evanescent state. This is the boundary-condition origin of the Type-II clouds.

\subsection{A General Criterion for Horizon-Decaying Clouds}

The above result also suggests a useful near-horizon criterion for when the same mechanism can appear in other magnetized rotating black holes. Consider a stationary and axisymmetric magnetized black-hole background with horizon generator
\begin{equation}
\chi = \partial_t + \Omega_H \partial_\varphi ,
\end{equation}
and horizon electrostatic potential
\begin{equation}
\Phi_H \equiv -\chi^\mu A_\mu|_{r=r_+}.
\end{equation}
For a charged scalar mode, with $m\in\mathbb{Z}$,
\begin{equation}
\Phi \sim e^{-i\omega t+im\varphi}S(\theta)R(r),
\end{equation}
the kinematic synchronization frequency is given by
\begin{equation}
\omega_c = m\Omega_H + q\Phi_H .
\end{equation}
Suppose that, after separation and a regular near-horizon field redefinition, the radial equation can be written near the horizon as
\begin{equation}
\frac{d^2Y}{dy^2} + k_h^2 Y = 0, \qquad
 k_h^2 = A_H(\omega-\omega_-)(\omega-\omega_+)
       = A_H\left[(\omega-\bar\omega_H)^2-\Delta_H^2\right],
\label{eq:general_gap}
\end{equation}
where $y\to -\infty$ at the horizon, $A_H>0$, and the roots are ordered as $\omega_-<\omega_+$. We define $\bar\omega_H=(\omega_-+\omega_+)/2$, $\Delta_H=(\omega_+-\omega_-)/2$, and $\mathcal{G}_H\equiv\Delta_H^2$. The two roots need not, in general, be centered exactly on the kinematic synchronization frequency $\omega_c$. The symmetric form used below for Kerr--BR is the special case $\bar\omega_H=\omega_c$ and $A_H=1$. In ordinary Kerr or Kerr-Newman backgrounds, the two roots coalesce at $\omega_-=\omega_+=\omega_c$, so there is no finite evanescent interval and the only zero-flux scalar cloud occurs at synchronization. By contrast, if a magnetized rotating black hole has
\begin{equation}
\Delta_H^2>0,
\label{eq:gap_condition}
\end{equation}
then the horizon wavenumber is imaginary in the finite interval
\begin{equation}
\omega_-<\omega<\omega_+.
\end{equation}
The two endpoints are the analogues of $\omega_1$ and $\omega_2$. Inside this interval the regular radial solution is
\begin{equation}
Y \sim e^{\kappa_H y}, \qquad \kappa_H=\sqrt{A_H}\sqrt{\Delta_H^2-(\omega-\bar\omega_H)^2},
\end{equation}
which decays toward the future event horizon. Since this boundary condition is a real exponential rather than a traveling wave, the same conserved radial current used in Eq.~\eqref{eq:horizon_flux} gives no net horizon transport. Therefore the superradiant energy exchange is quenched not by kinematic synchronization alone, but by the conversion of the horizon channel from propagating to evanescent. If the same mode is also confined at infinity, the usual bound-state quantization then yields a stationary horizon-decaying scalar cloud.

In the Kerr-BR weak-field sector studied here, Eq.~\eqref{eq:general_gap} is realized with
\begin{equation}
\mathcal{G}_H=\Delta_H^2=\omega_c^2-D
=\frac{a^2B^2q^2r_+^2}{16M^2}.
\end{equation}
Thus the gap vanishes whenever the rotation, magnetic field, or scalar charge is removed, and the Type-II branch collapses back to the usual synchronized-cloud picture. This shows that the horizon-decaying clouds found below are not an artifact of notation, but a concrete realization of a broader criterion: a positive rotation--magnetic horizon gap is sufficient to generate an evanescent scalar-cloud branch.

\paragraph{Independent realization: Melvin--Kerr.}

The same root-splitting structure is visible in the weak-field separable sector of the Melvin--Kerr geometry. For a charged scalar, the near-horizon radial equation obtained in that setting takes the form
\begin{equation}
\begin{aligned}
\frac{d^2 Y}{dy^2}+k_{\rm MK}^2 Y&=0, \\
k_{\rm MK}^2&=(\omega-\omega_{\rm syn}^{\rm MK})^2
 +\alpha_{\rm MK}(\omega-\omega_{\rm syn}^{\rm MK}),
\end{aligned}
\label{eq:melvin_kerr_gap}
\end{equation}
where
\begin{equation}
\alpha_{\rm MK}=\frac{8bM^2q}{a}.
\label{eq:melvin_kerr_alpha}
\end{equation}
Here $b$ denotes the Melvin magnetic-field parameter, and $\omega_{\rm syn}^{\rm MK}=ma/(2Mr_+)$ is the horizon synchronization frequency in the gauge $A_t(\infty)=0$ used in the Melvin--Kerr analysis. Equation~\eqref{eq:melvin_kerr_gap} is reported in the weak-field regime $bM\ll1$, where the charged Klein--Gordon equation is separable \cite{siahaan2025melvinkerrclouds,siahaan2025melvinkerrsuperradiance}. Completing the square gives
\begin{equation}
k_{\rm MK}^2=\left(\omega-\omega_{\rm syn}^{\rm MK}+\frac{\alpha_{\rm MK}}{2}\right)^2
-\frac{\alpha_{\rm MK}^2}{4},
\end{equation}
so that the two horizon roots are
\begin{equation}
\omega_-^{\rm MK}=\omega_{\rm syn}^{\rm MK}-\alpha_{\rm MK},
\qquad
\omega_+^{\rm MK}=\omega_{\rm syn}^{\rm MK}.
\label{eq:melvin_kerr_roots}
\end{equation}
For $bq/a>0$, the interval
\begin{equation}
\omega_{\rm syn}^{\rm MK}-\frac{8bM^2q}{a}<\omega<\omega_{\rm syn}^{\rm MK}
\end{equation}
has $k_{\rm MK}^2<0$. Thus Melvin--Kerr supplies an independent example in which the propagating horizon channel terminates below the synchronization frequency and is replaced by an evanescent channel. In the notation of Eq.~\eqref{eq:general_gap}, its positive gap is $\Delta_{\rm MK}^2=\alpha_{\rm MK}^2/4$, while the center of the two roots is shifted away from $\omega_{\rm syn}^{\rm MK}$. The Kerr--BR result is therefore not an isolated algebraic coincidence: it is the symmetric realization of a more general two-root near-horizon dispersion pattern. We emphasize, however, that the Melvin--Kerr observation is presently a near-horizon diagnostic in the weak-field separable treatment; establishing a complete Type-II cloud spectrum in the fully backreacted, nonseparable geometry would require a separate analysis.

For the Kerr--BR realization studied in this work, the general roots $\omega_-$ and $\omega_+$ reduce to the transition frequencies $\omega_1$ and $\omega_2$ defined above, while the quantity $D$ specifies their location through the Kerr--BR near-horizon potential. Returning to this geometry-specific notation, the sign of $k_h^2$ is controlled by the joint appearance of rotation and magnetic coupling in $\omega_c$ and $D$. The near-horizon boundary condition therefore bifurcates into two distinct branches:
\begin{itemize}
    \item \textbf{Propagating Branch ($k_h^2 > 0$):} The wave is physically ingoing, $Y \sim e^{i k_h y}$ for $0 < \omega < \omega_1$, permitting rotational energy extraction (superradiance).
    \item \textbf{Horizon-Evanescent Branch ($k_h^2 < 0$):} The horizon wavenumber becomes imaginary, $k_h = -i\kappa_h$. The regular solution exponentially decays toward the horizon, $Y \sim e^{\kappa_h y}$. This is the branch that gives rise to Type-II scalar clouds.
\end{itemize}

This comparison suggests a broader physical interpretation. The horizon-evanescent branch should not be regarded as tied to the particular Kerr--BR asymptotics, but as a mechanism to be tested across magnetized rotating geometries. In realistic environments, accretion flows, jets, and surrounding plasma generate magnetic configurations that are neither exactly uniform nor strictly Melvin \cite{blandford1977electromagnetic,wald1974black}. Whether their near-horizon expansions produce a nonzero root separation, and whether the resulting local evanescent branch can be matched to global confinement, are open but concrete questions for future work. The present criterion therefore provides a diagnostic for realistic magnetic field geometries rather than a claim that every astrophysical magnetosphere must support such clouds.

\subsection{Classification of Bound States}

Defining the asymptotic cutoff frequency as $\omega_f = -a B q + \sqrt{B m q + a^2 B^2 q^2 + \mu_s^2}$, the bound states naturally divide into three regimes:

\begin{enumerate}
    \item \textbf{Superradiant bound states} $\left( 0 < \omega < \min\{\omega_1, \omega_f\} \right)$: $k_h^2 > 0$. The wave is propagating at the horizon and extracts rotational energy.
    \item \textbf{Horizon-decaying bound states} $\left( \omega_1 \le \omega \le \min\{\omega_2, \omega_f\} \right)$: $k_h^2 \le 0$. The joint rotational and magnetic couplings convert the horizon behavior into an evanescent decay, $Y \sim e^{\kappa_h y}$, quenching the superradiant flux.
    \item \textbf{Non-superradiant bound states} $\left( \omega_2 < \omega < \omega_f \right)$: $k_h^2 > 0$ but $\omega > \omega_c$. The wave is again propagating at the horizon, now in the absorbing regime.
\end{enumerate}

\section{Analytic Bound-State Spectrum and Rotation--Magnetic Quenching}
\label{sec:analytic_quenching}

\begin{figure}[htbp]
    \centering
    \includegraphics[width=0.45\textwidth]{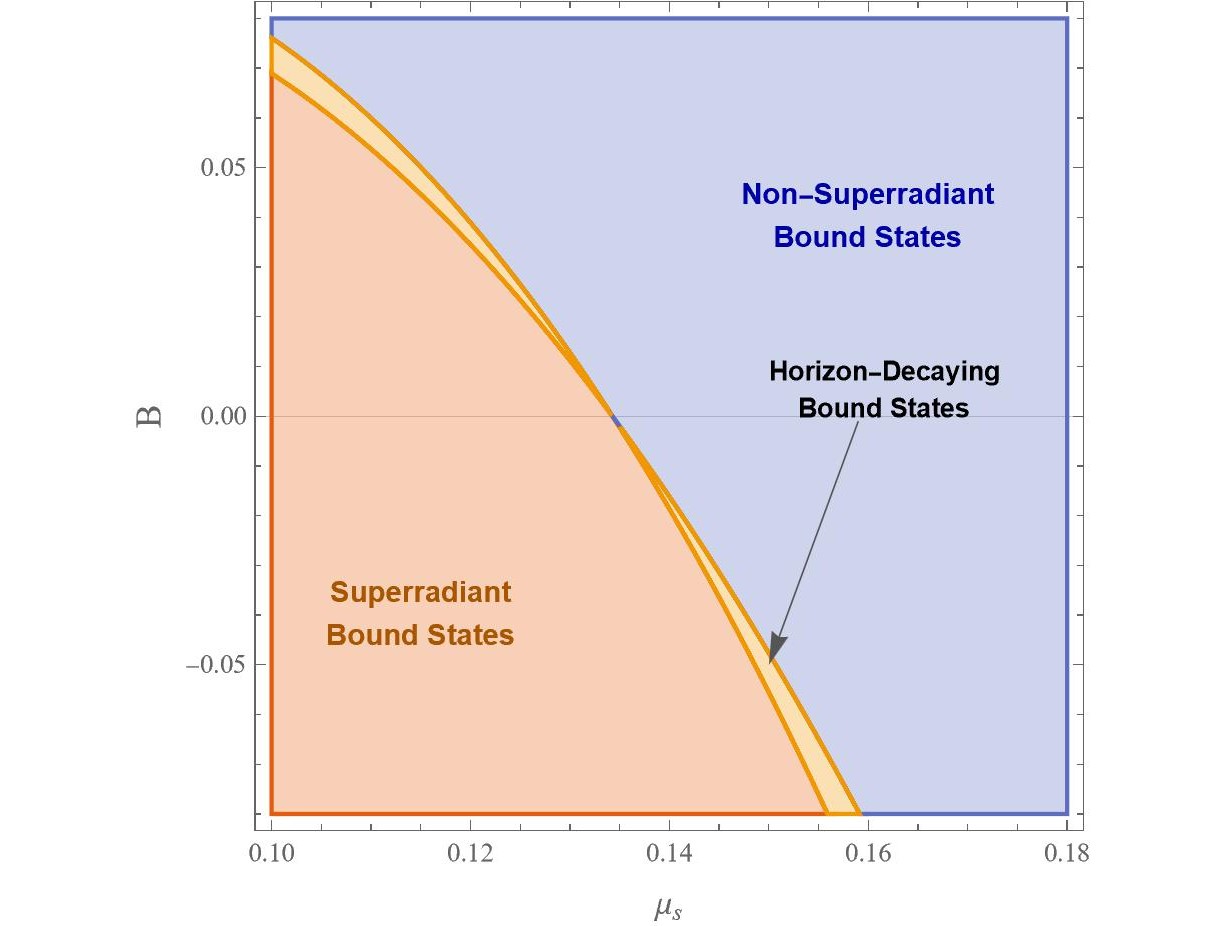} \hfill
    \includegraphics[width=0.45\textwidth]{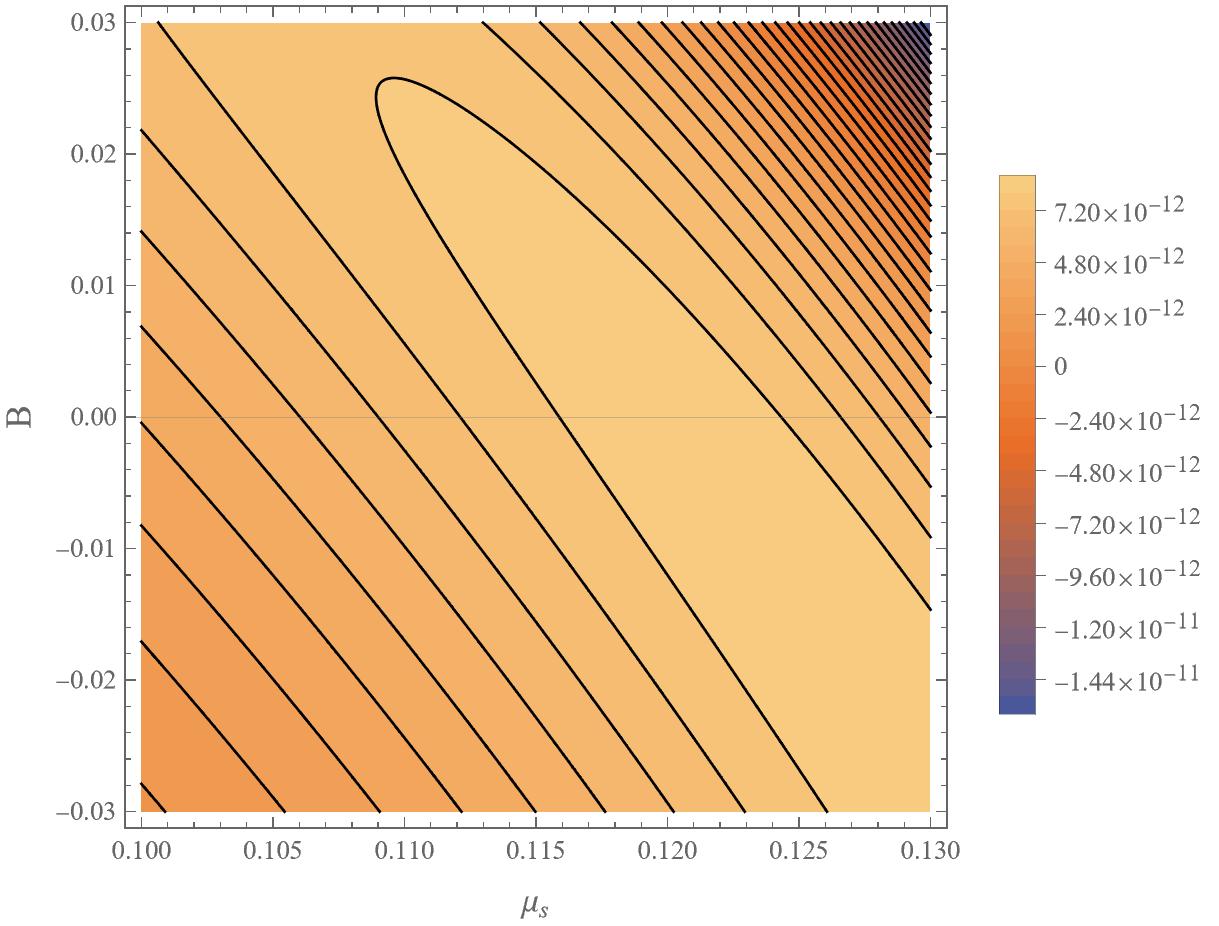}
    \caption{Left: Parametric phase diagram in the $(\mu_s, B)$ plane highlighting three distinct bound-state regimes for $M=1$, $a=0.5$, $q=0.1$, $n=0$, and $l=m=1$.
The colored regions correspond to Superradiant Bound States (orange), Non-Superradiant Bound States (blue), and the critical Horizon-Decaying Bound States (gold).
Right: Contour visualization of the magnitude of instability growth rate $\gamma = \text{Im}(\delta\omega)$ within the superradiant bound state regime.
This panel is directly generated by visualizing the analytical formula, rather than from full numerical integrations, explicitly mapping the continuous transition towards the critical quenching threshold ($\gamma=0$).}
    \label{fig:analytical_gamma}
\end{figure}

The transition from a propagating to an exponentially decaying horizon boundary condition changes the near-horizon character of the bound states. Employing the matched asymptotic expansion method \cite{detweiler1980klein}, we analytically map the complex frequency. The weak-$B$ truncation used to obtain the separated radial equation and the matched weak-binding expansion used here are distinct approximations. In the weak-binding regime $|k|M\ll1$, the near and far solutions overlap in $M\ll r-r_+\ll|k|^{-1}$. Appendix~\ref{app:matching} contains the full matching; here we retain only the relations needed for the spectrum and the quenching mechanism.

In the far region, $r-r_+\gg M$, the radial equation reduces to a Coulomb-type bound-state equation. With
\begin{equation}
k^2=\omega^2-\mu_s^2-Bmq+2aBq\omega ,
\qquad
\nu=-\frac{i\left(BmMq+M\mu_s^2-3aBMq\omega-2M\omega^2\right)}{k},
\end{equation}
the solution regular at spatial infinity can be written schematically as
\begin{equation}
R_{\rm far}=e^{ix/2}x^l
U\!\left(1+l+\epsilon^2-\nu,2(1+l+\epsilon^2),-ix\right),
\qquad x=2kr ,
\label{eq:main_far_solution}
\end{equation}
where $U$ is the confluent hypergeometric function. Here
$\epsilon^2$ denotes the subleading correction to the far-region centrifugal
coefficient; it is retained as a bookkeeping term but omitted from the leading
spectrum. This solution decays at infinity when $k=i|k|$, which is precisely the bound-state condition.

In the near region, $r-r_+\ll |k|^{-1}$, we introduce
$z=(r-r_+)/(r_+-r_-)$. The solution used for matching is
\begin{equation}
R_{\rm near}=\left(\frac{z}{1+z}\right)^{i\alpha}
F(-l,l+1;1+2i\alpha;-z).
\label{eq:main_near_solution}
\end{equation}
Here $F$ is the Gauss hypergeometric function and $\alpha$ is a dimensionless
Frobenius index. Near the horizon, the two independent local behaviors are
$R\sim z^{\pm i\alpha}$; Eq.~\eqref{eq:main_near_solution} selects the
$z^{i\alpha}$ branch. Within the retained weak-field radial equation, $\alpha$
is related to the horizon wavenumber by
\begin{equation}
\alpha^2=\left(\frac{2Mr_+}{r_+-r_-}\right)^2 k_h^2 .
\label{eq:alpha_kh}
\end{equation}
Thus $\alpha$ is real on the propagating branch. On the horizon-evanescent
branch, where $k_h=-i\kappa_h$, regularity fixes
\[
\alpha=-i\beta,
\qquad
\beta\equiv\frac{2Mr_+}{r_+-r_-}\kappa_h>0,
\]
so that the selected solution behaves as $R_{\rm near}\sim z^{\beta}$ and
vanishes as $z\to0$. The explicit expression for $\alpha^2$ in terms of
$(M,a,B,q,m,\omega)$ is given in Appendix~\ref{app:matching}.

In the overlap region, the small-$|x|$ far solution and the
large-$z$ near solution share the behaviors $r^l$ and $r^{-l-1}$. Matching
their coefficients gives the resonance condition (see
Appendix~\ref{app:matching})
\begin{equation}
    \frac{\Gamma(-l - \nu - \epsilon^2) \Gamma(2l + 2)}
    {\Gamma(l - \nu + 1 + \epsilon^2) \Gamma(-2l - 2\epsilon^2)}
    = -2\alpha \left[ 2k(r_+ - r_-) \right]^{2l+1}
    \prod_{j=1}^l (j^2 + 4\alpha^2)
    \frac{\Gamma(l+1) \Gamma(-2l-1)}
    {\Gamma(2l+1)\Gamma(-l)} .
\label{eq:main_match}
\end{equation}
At fixed $\alpha$, the right-hand side is of order
$[|k|(r_+-r_-)]^{2l+1}$, so its leading-order zero gives the pole condition
\begin{equation}
l-\nu^{(0)}+1+\epsilon^2=-n,\qquad n=0,1,2,\ldots ,
\end{equation}
which gives the leading-order real frequency $\omega^{(0)}$ and
the asymptotic wavenumber $k^{(0)}$:
\begin{align}
\omega^{(0)} &= \omega_f - \frac{M^2 \left[ \omega_f^2 - \left( a^2 B^2 q^2 - a B q \sqrt{a^2 B^2 q^2+B m q+\mu_s^2} \right) \right]^2}{2 (l+n+1)^2 \sqrt{a^2 B^2 q^2+B m q+{\mu_s}^2}}, \\
k^{(0)} &= i \sqrt{\frac{M^2 \sqrt{a^2 B^2 q^2+B m q+{\mu_s}^2 } \, {\omega_f}^3}{(l+n+1)^2}}.
\end{align}
Thus $k^{(0)}$ is imaginary for the bound state. The first
matching correction follows by expanding the quantization condition about this
pole:
\begin{equation}
\nu=\nu^{(0)}+\delta\nu,
\qquad
\omega=\omega^{(0)}+\delta\omega,
\qquad
|\delta\omega|\ll|\omega^{(0)}| .
\label{eq:frequency_expansion}
\end{equation}
Equation~(25) defines $k(\omega)$ and $\nu(\omega)$; it is not itself an
expansion. The object expanded is the quantization condition about its leading
pole, equivalently $\nu(\omega)$ about $\nu^{(0)}$, rather than the magnetic
field. At fixed $\alpha^{(0)}$,
$\delta\nu=O\!\left([|k^{(0)}|(r_+-r_-)]^{2l+1}\right)$. The detailed pole
expansion is given in Appendix~\ref{app:matching}.

The perturbative frequency shift $\delta\omega$ is related to
$\delta\nu$ through
\begin{equation}
\delta\nu =
\left.\frac{\partial\nu}{\partial\omega}\right|_{\omega=\omega^{(0)}}\delta\omega ,
\end{equation}
where the variation $\delta\nu$ is
\begin{equation}
\delta \nu = 2\alpha^{(0)} \left[ 2k^{(0)}(r_+ - r_-) \right]^{2l+1} \frac{(2l+1+n)!}{n!} \left[ \frac{l!}{(2l)!(2l+1)!} \right]^2 \prod_{j=1}^l \left( j^2 + 4{\alpha^{(0)}}^2 \right) \label{deltanu}.
\end{equation}
Here $\alpha^{(0)}\equiv\alpha(\omega^{(0)})$ is evaluated at the
leading-order frequency. The growth or decay rate
$\gamma=\operatorname{Im}(\delta\omega)$ is determined by $\alpha^{(0)}$ and
$k^{(0)}$.

In conventional superradiant scenarios (the orange regime in Fig.~\ref{fig:analytical_gamma}), $\alpha^{(0)}$ is real. Since
\begin{equation}
\left[k^{(0)}\right]^{2l+1}=i(-1)^l |k^{(0)}|^{2l+1},
\end{equation}
Eq.~\eqref{deltanu} is imaginary at leading order, and hence $\delta\omega$ has a nonzero imaginary part. The right panel of Fig.~\ref{fig:analytical_gamma} displays the magnitude of this growth rate, obtained by directly visualizing the analytical formula in the superradiant bound-state regime.

As explicitly mapped by the contours, the growth rate $\gamma$ exhibits a coupled dependence on both the scalar mass $\mu_s$ and the magnetic field strength $B$. The parameter space features a distinct ``ridge'' of maximum instability, representing the optimal balance between superradiant energy extraction and wave confinement. For a fixed magnetic field $B$, increasing the scalar mass $\mu_s$ changes the effective potential well and progressively suppresses the amplification efficiency. Conversely, tuning $B$ modulates both the asymptotic confinement and the near-horizon electromagnetic coupling. Moving away from the ridge, either by shifting $\mu_s$ or by adjusting $B$ toward the critical transition boundary, drives the growth rate down to zero.

When the system enters the horizon-decaying regime (the gold band in the left
panel of Fig.~\ref{fig:analytical_gamma}), the physical branch is
$\alpha^{(0)}=-i|\alpha^{(0)}|$. Since $k^{(0)}=i|k^{(0)}|$, the combination
$\alpha^{(0)}[k^{(0)}]^{2l+1}$ in Eq.~\eqref{deltanu} is real. The remaining
factors are also real, and hence
\begin{equation}
    \text{Im}(\delta\omega) = \gamma = 0. \label{eq:gamma_zero}
\end{equation}
This is the analytic quenching mechanism: the magnetic-rotation coupling that
turns the horizon wave into an evanescent mode also removes the leading
dissipative imaginary part of the bound-state frequency. Appendix~\ref{app:matching}
gives the explicit phase cancellation.

\section{Numerical Evidence for the Two Branches}
\label{sec:numerical}

To validate the analytical framework and explore the detailed spatial profiles, we solve the Schr\"{o}dinger-like radial equation \eqref{eq:yreq} numerically via a direct integration method coupled with a shooting algorithm \cite{press2007numerical}. The detailed numerical strategy is outlined in Appendix \ref{app:numerical}. We focus on the fundamental and overtone modes ($l=m=1$) with fixed parameters $M=1, a=0.5$, and $q=0.3$ to plot the wavefunctions as $R(y)$.

\subsection{Propagating Reference Clouds (Type-I)}

\begin{figure}[htbp]
    \centering
    \includegraphics[width=0.45\textwidth]{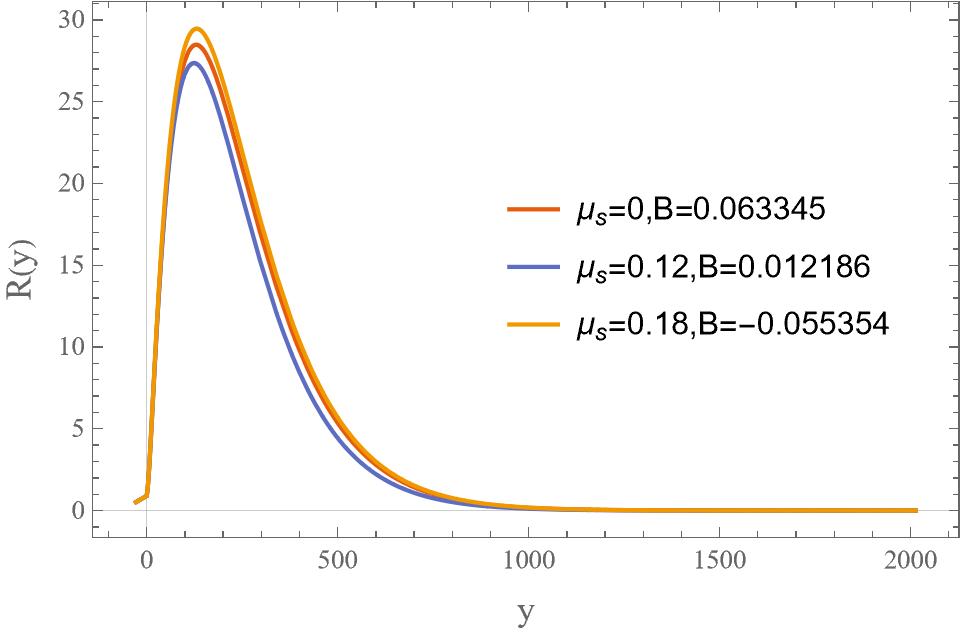} \hfill
    \includegraphics[width=0.45\textwidth]{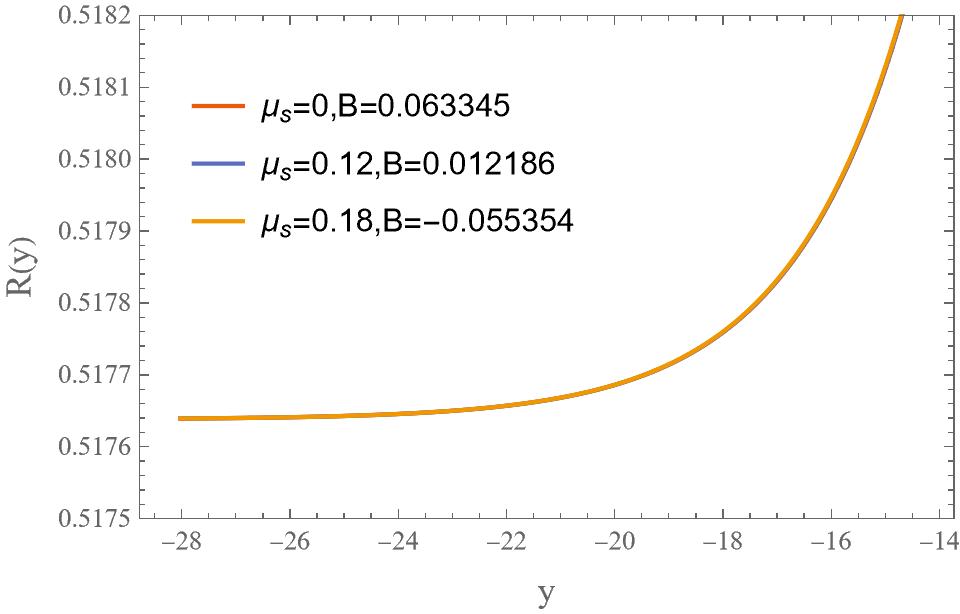} \\
    \vspace{0.15cm}
    \includegraphics[width=0.45\textwidth]{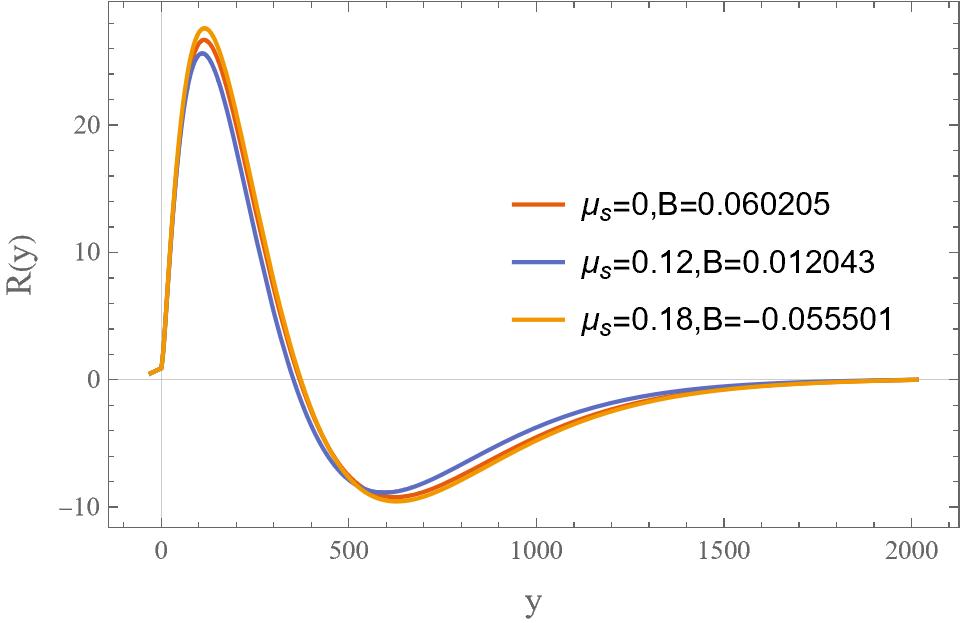} \hfill
    \includegraphics[width=0.45\textwidth]{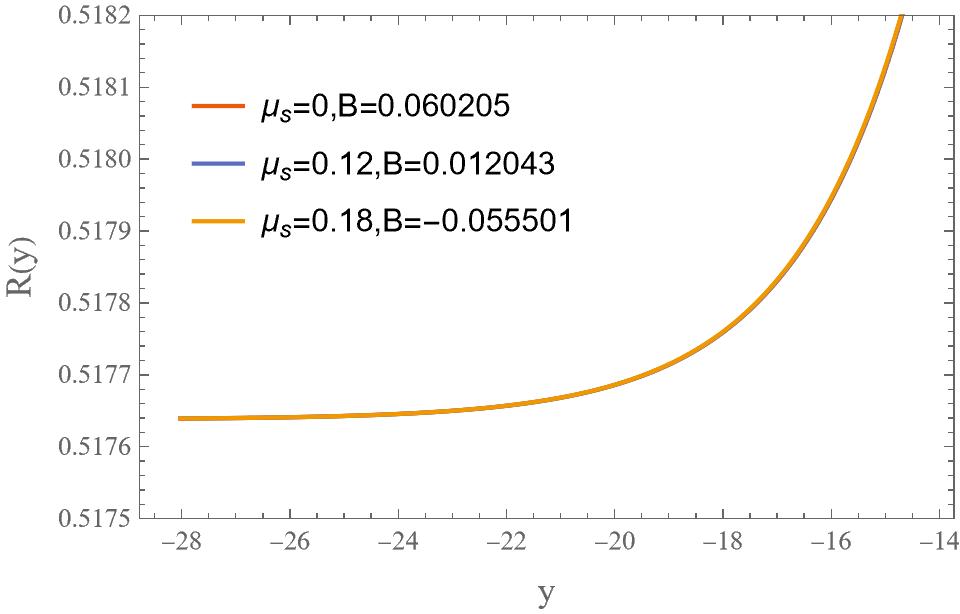} \\
    \vspace{0.15cm}
    \includegraphics[width=0.45\textwidth]{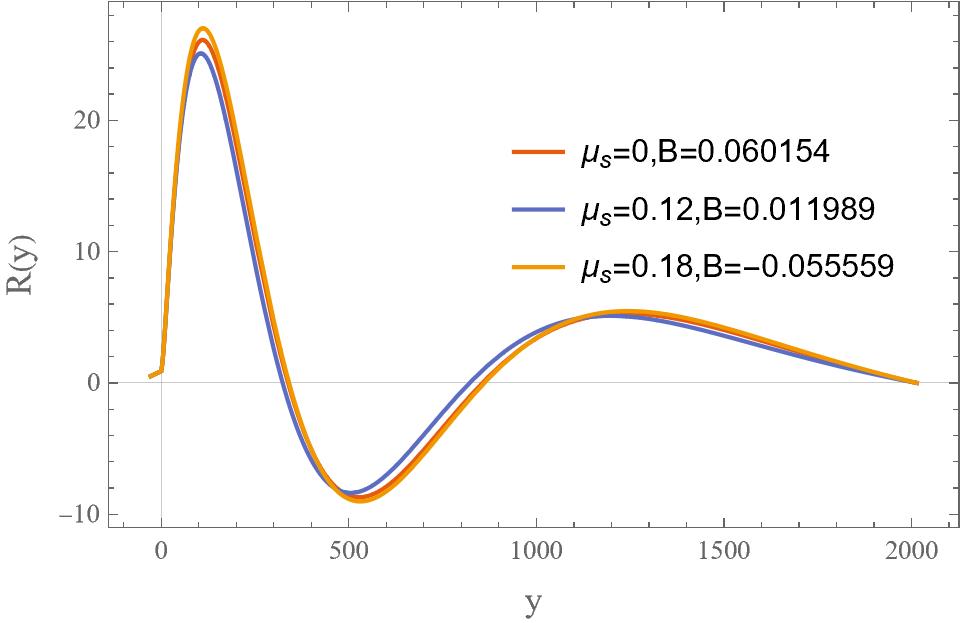} \hfill
    \includegraphics[width=0.45\textwidth]{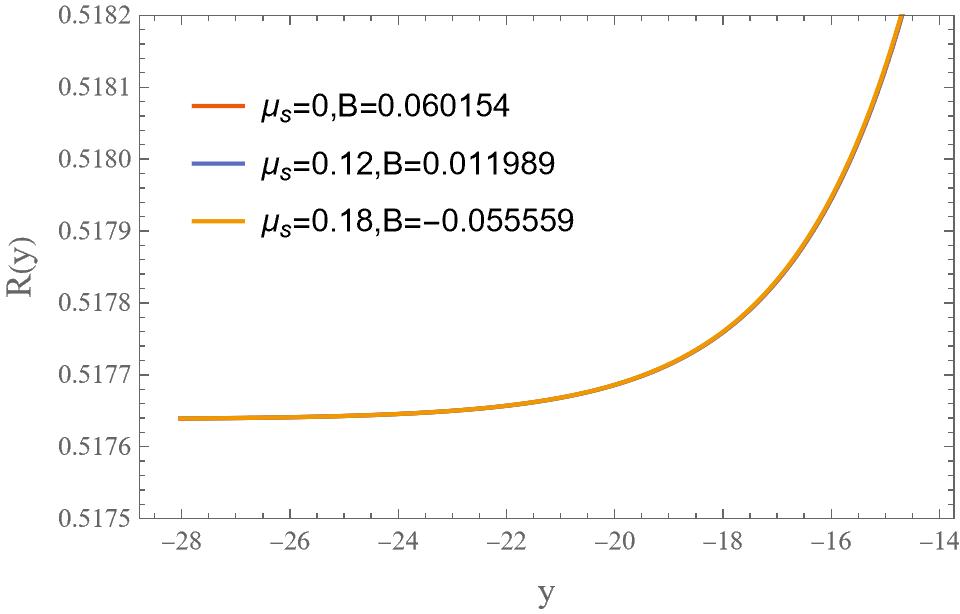}
    \caption{Radial wavefunctions $R(y)$ of the propagating Type-I reference clouds ($n=0, 1, 2$) evaluated at $\omega=\omega_1$. Left panels: Global bound-state decay at spatial infinity. Right panels: Near-horizon behavior showing a non-zero constant amplitude, characteristic of the zero-wavenumber edge of the propagating branch.}
    \label{fig:classical_clouds_n012}
\end{figure}

Figure~\ref{fig:classical_clouds_n012} illustrates the radial wavefunctions of the propagating Type-I clouds, which serve as the reference branch for comparison with the horizon-decaying solutions. By tuning the frequency exactly to the lower transition point $\omega = \omega_1$, the system sits at the boundary between propagating superradiant modes and the magnetically induced evanescent band. As shown in the left panels, the scalar field decays exponentially at spatial infinity, forming a localized bound state. Notably, the $\mu_s=0$ curves confirm that even for a massless scalar field, the external magnetic field provides an effective mass squared ($\mu_{\text{eff}}^2 = B m q - 2 a B q \omega$) at spatial infinity, acting as a potential barrier to support the bound state. 

The right panels show that as $y \to -\infty$ (approaching the horizon), the amplitude of all wavefunctions plateaus to a non-zero constant. Altering the scalar mass $\mu_s$ and the magnetic field $B$ does not change this qualitative near-horizon structure. They all asymptotically approach a constant value, confirming that Type-I clouds are the zero-wavenumber edge of the propagating branch rather than the horizon-evanescent branch.

\begin{figure}[htbp]
    \centering
    \includegraphics[width=0.45\textwidth]{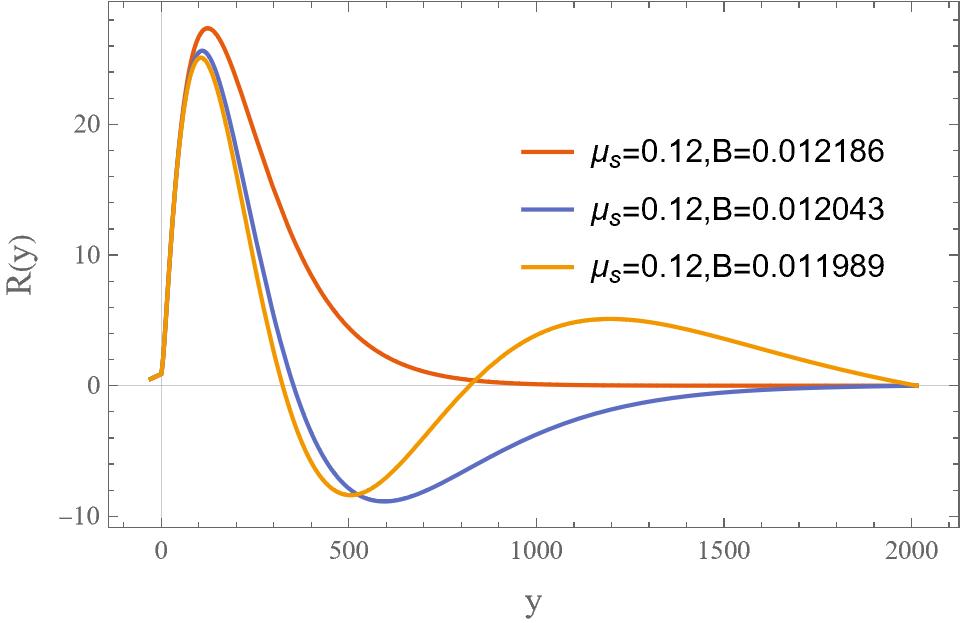} \hfill
    \includegraphics[width=0.45\textwidth]{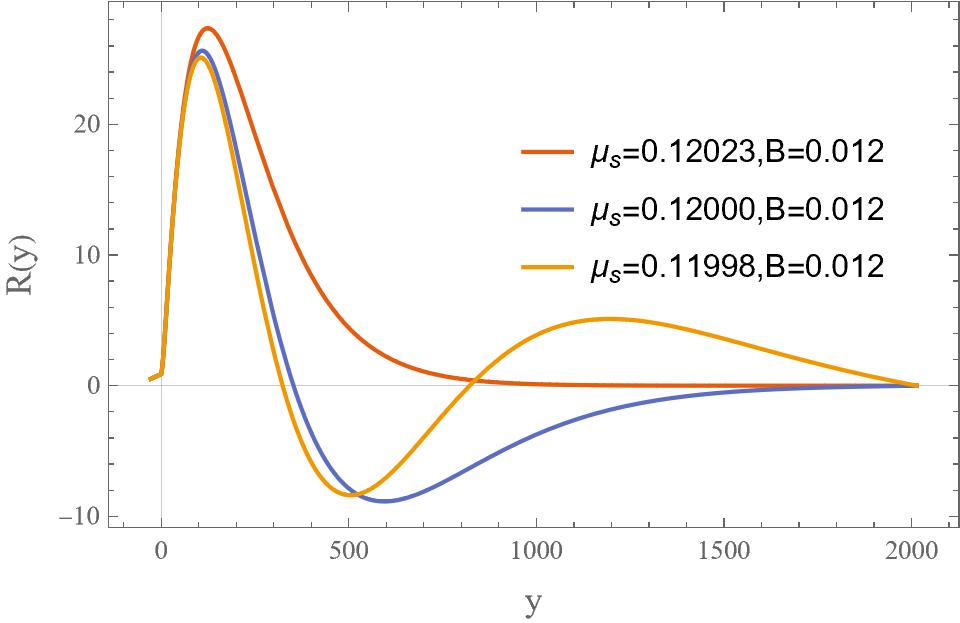} 
    \caption{Comparison of different overtone numbers ($n=0, 1, 2$) for Type-I propagating reference clouds. Left: Fixed scalar mass $\mu_s \approx 0.12$. Right: Fixed magnetic field $B = 0.012$. The counterpart parameter must be finely tuned to maintain the synchronized bound state and excite higher nodes.}
    \label{fig:compare_overtones_type1}
\end{figure}

To further investigate the excitation spectrum, we compare the fundamental mode ($n=0$) with the overtones ($n=1, 2$). The overtone number $n$ corresponds to the number of radial nodes in the wavefunction. As shown in Fig.~\ref{fig:compare_overtones_type1}, accommodating higher overtones requires precise adjustments to the potential well. The left panel shows that when the scalar mass $\mu_s$ is fixed, gradually decreasing the magnetic field $B$ relaxes the spatial confinement, allowing the wavefunction to extend further and excite more radial nodes. Conversely, the right panel demonstrates that when $B$ is fixed, slightly decreasing $\mu_s$ achieves a similar spatial extension and overtone excitation.

\begin{figure}[htbp]
    \centering
    \includegraphics[width=0.45\textwidth]{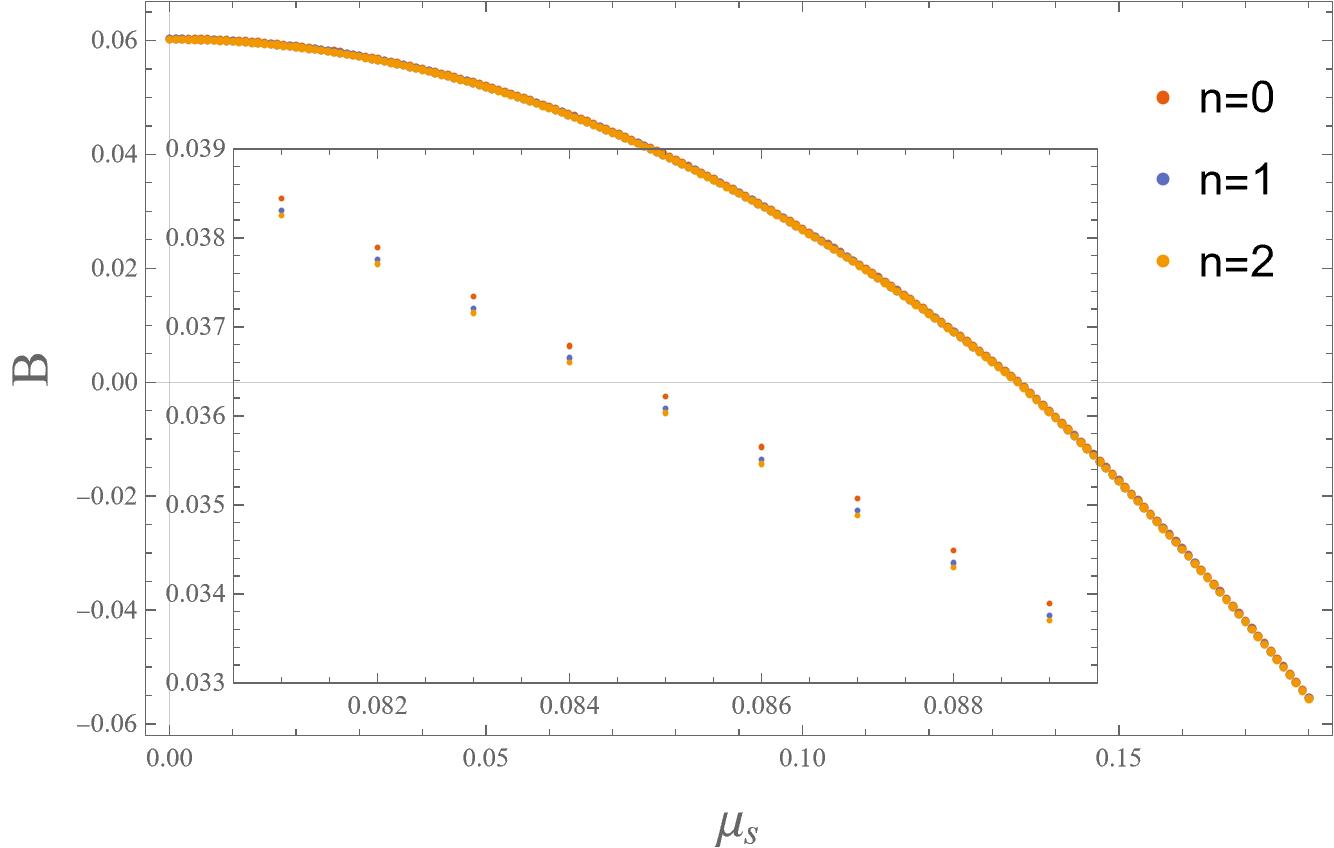} \hfill
    \includegraphics[width=0.45\textwidth]{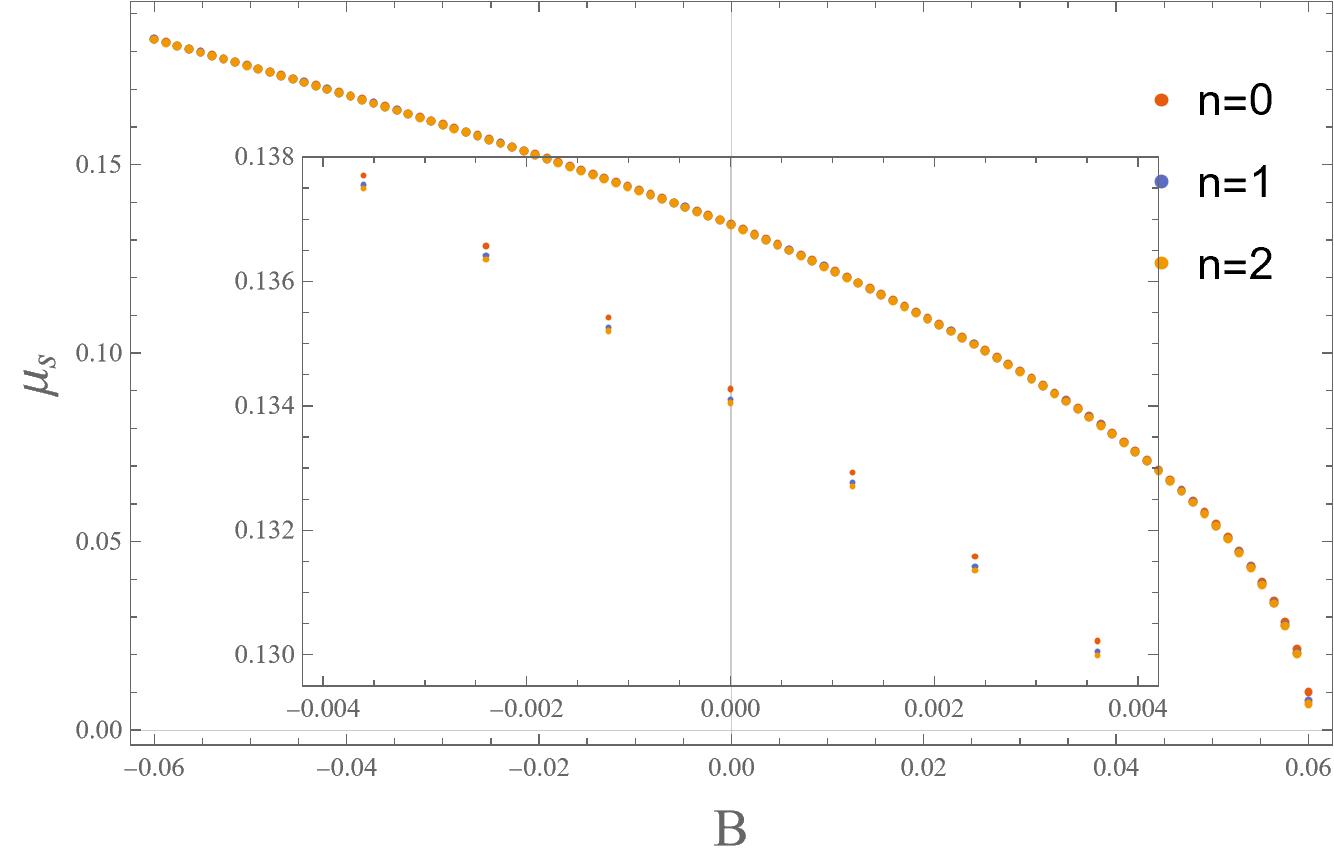} \\
    \vspace{0.2cm}
    \includegraphics[width=0.45\textwidth]{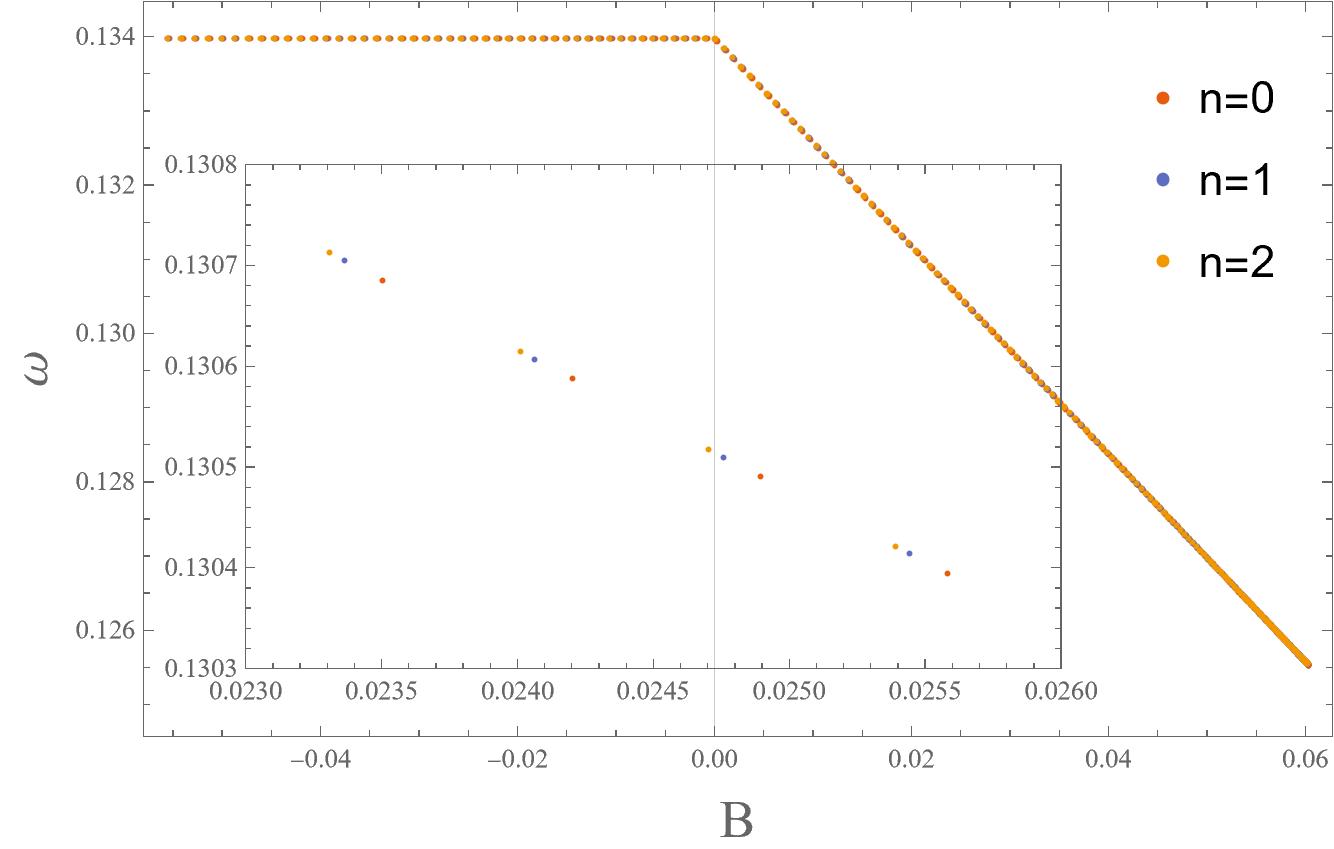} \hfill
    \includegraphics[width=0.45\textwidth]{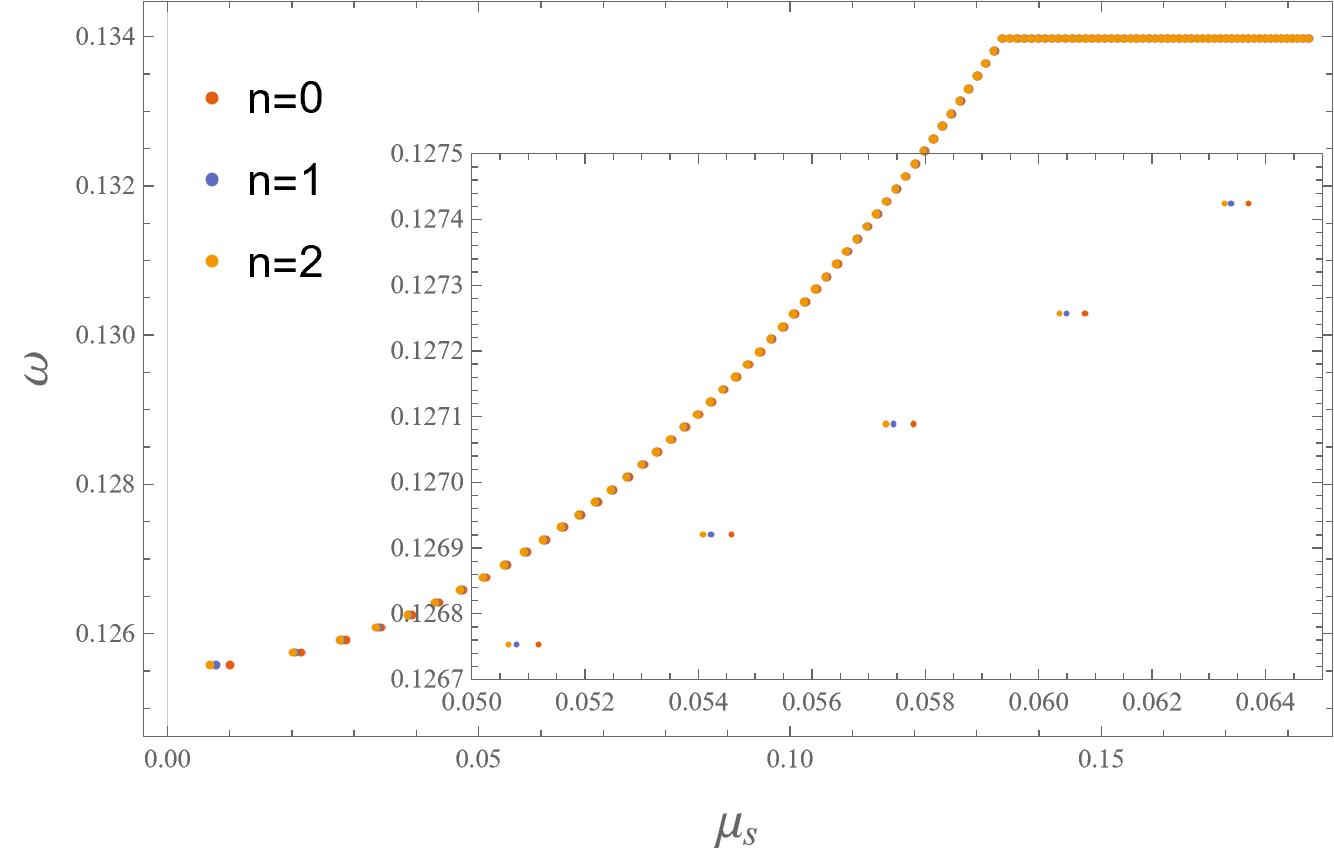}
    \caption{The permissible parameter space for propagating Type-I reference clouds at $\omega=\omega_1$. The upper panels show the bounds on $\mu_s$ and $B$, while the lower panels display the evolution of $\omega_1$. The embedded insets resolve the discrete fine-structure splitting induced by the overtone number $n$.}
    \label{fig:type1_parameter_space}
\end{figure}

Figure~\ref{fig:type1_parameter_space} maps the permissible parameter space and the corresponding critical frequency $\omega_1$ for these Type-I propagating reference clouds. The upper panels display the $(\mu_s, B)$ space slices. The left panel shows the required $B$ values for different modes ($n=0,1,2$) at fixed $\mu_s$, while the right panel shows the corresponding $\mu_s$ values at fixed $B$. The embedded insets highlight the discrete fine-structure splitting between different overtones. Furthermore, these panels reveal that to satisfy the physical condition $\mu_s > 0$, the magnetic field is strictly bounded by $B < 0.063345$. The lower panels illustrate how the synchronized frequency $\omega_1$ evolves with $B$ (left) and $\mu_s$ (right), again demonstrating the discrete frequency splitting for different radial nodes in the insets.

\subsection{Horizon-Decaying Scalar Clouds from Rotation--Magnetic Coupling (Type-II)}

\begin{figure}[htbp]
    \centering
    \includegraphics[width=0.45\textwidth]{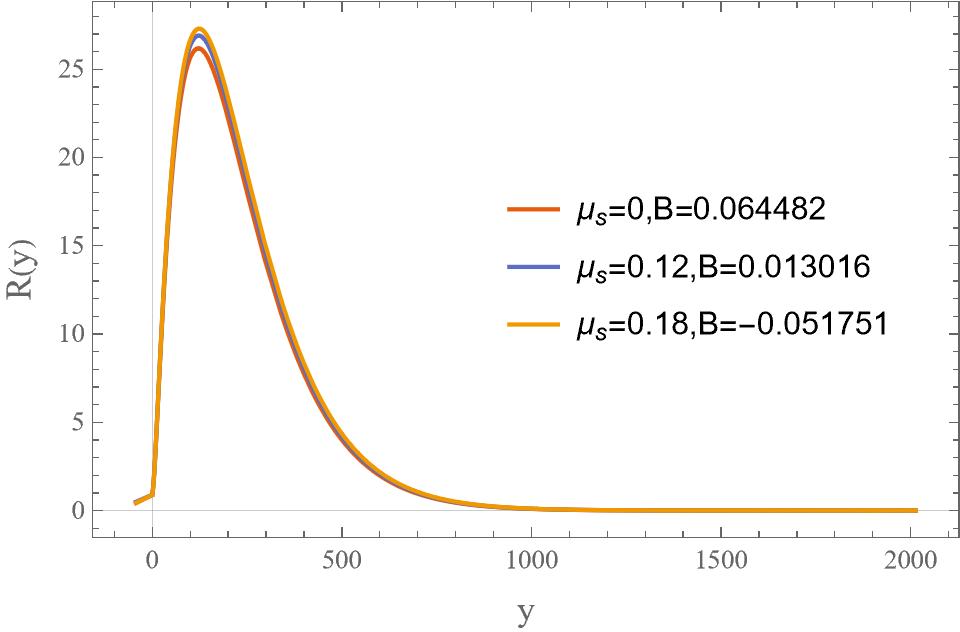} \hfill
    \includegraphics[width=0.45\textwidth]{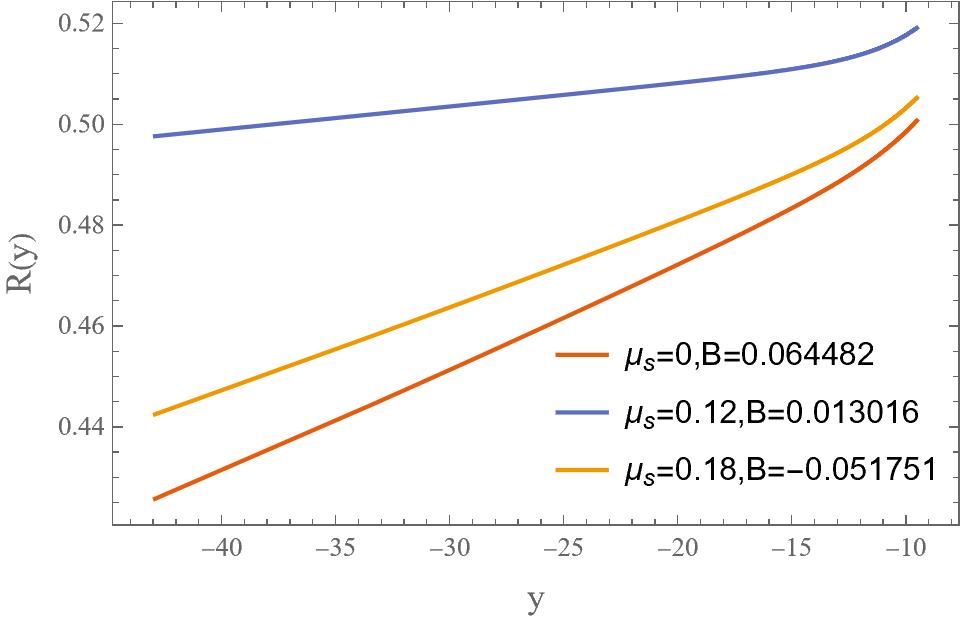} \\
    \vspace{0.15cm}
    \includegraphics[width=0.45\textwidth]{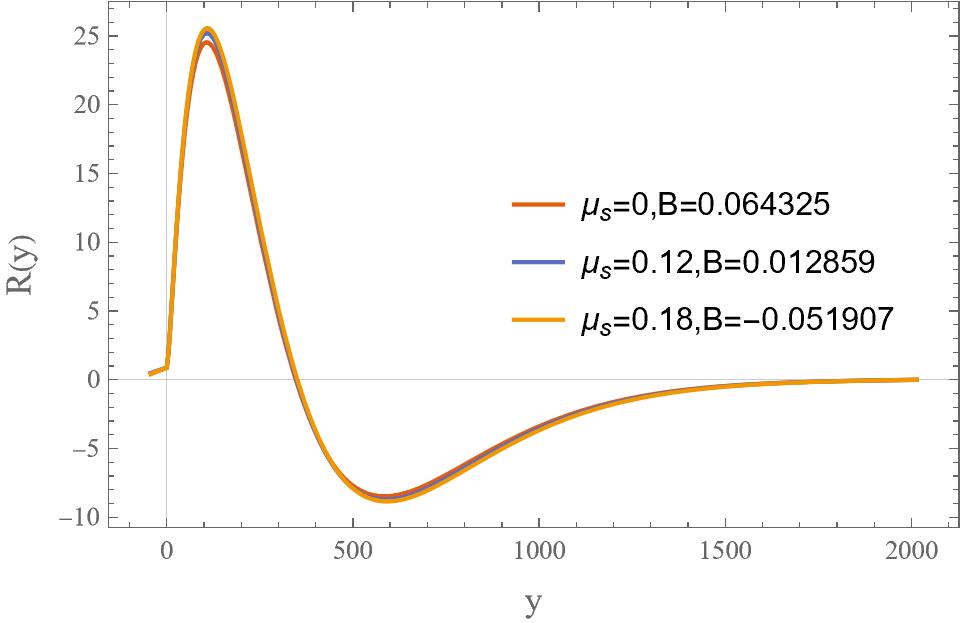} \hfill
    \includegraphics[width=0.45\textwidth]{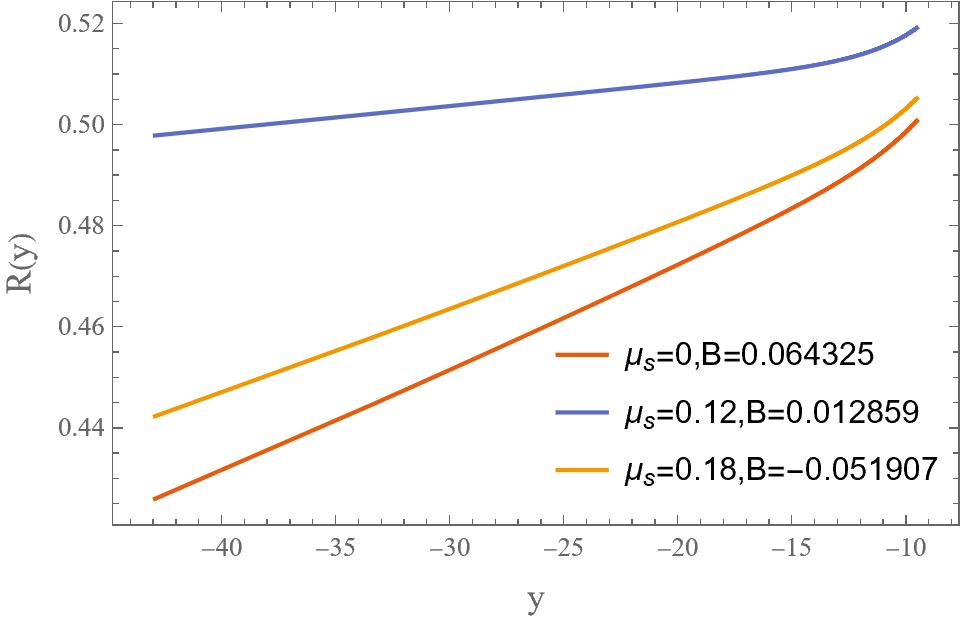} \\
    \vspace{0.15cm}
    \includegraphics[width=0.45\textwidth]{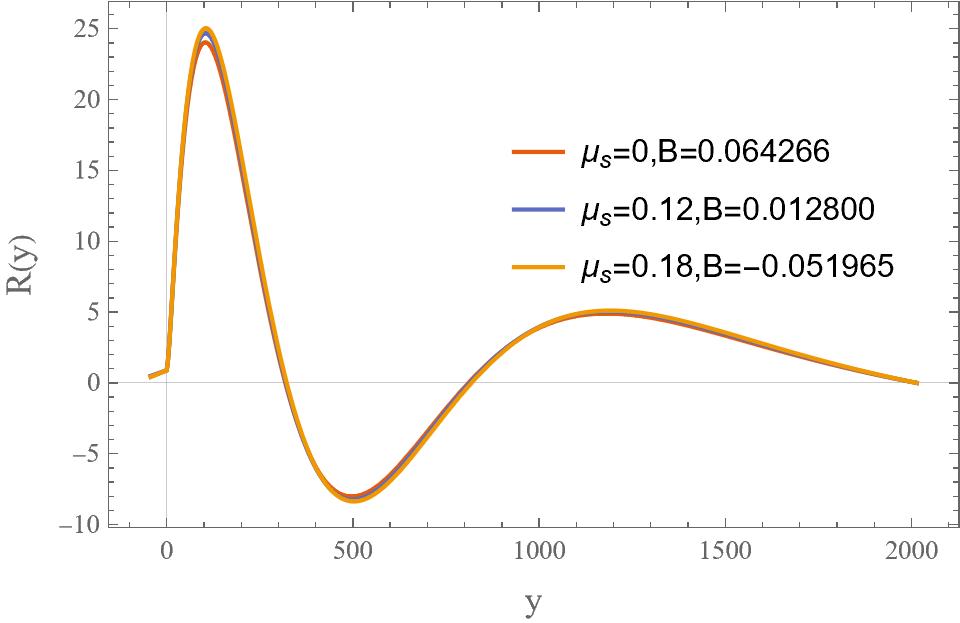} \hfill
    \includegraphics[width=0.45\textwidth]{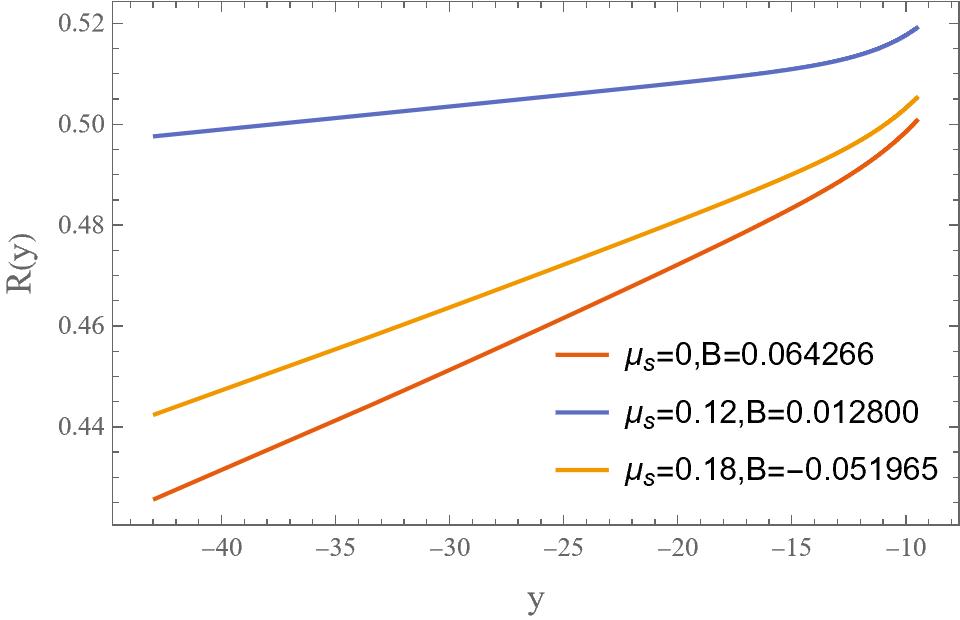}
    \caption{Radial wavefunctions $R(y)$ of the horizon-decaying scalar clouds ($n=0, 1, 2$) at $\omega=\omega_c$. The right panels illustrate the exponential decay condition at the event horizon and show that varying parameters directly affect the steepness of the decay.}
    \label{fig:evanescent_clouds_n012}
\end{figure}

Figure~\ref{fig:evanescent_clouds_n012} depicts the Type-II horizon-decaying scalar clouds evaluated at $\omega = \omega_c$, which lies inside the rotation--magnetic quenching band. These solutions are the main horizon-decaying configurations studied here. Similar to Type-I clouds, the $\mu_s=0$ curves verify that the effective scalar mass provided by the external magnetic field is sufficient to support these bound states at infinity. 

The right panels reveal the key physical difference near the event horizon. Unlike the constant plateau of Type-I clouds, the amplitude of Type-II wavefunctions exponentially decays toward zero as $y \to -\infty$, confirming the $Y \sim e^{\kappa_h y}$ boundary condition. This behavior is not obtained by merely tuning to a synchronization point; it is produced when the magnetic coupling and black-hole rotation jointly drive $k_h^2$ negative. Because the magnitude of the decay constant $\kappa_h$ is small in the examples shown, this exponential decay is strongly stretched and appears visually as a gradual linear growth extending outward from the horizon. Changing $\mu_s$ and $B$ directly alters $\kappa_h$, resulting in different degrees of exponential decay at the horizon.

\begin{figure}[htbp]
    \centering
    \includegraphics[width=0.45\textwidth]{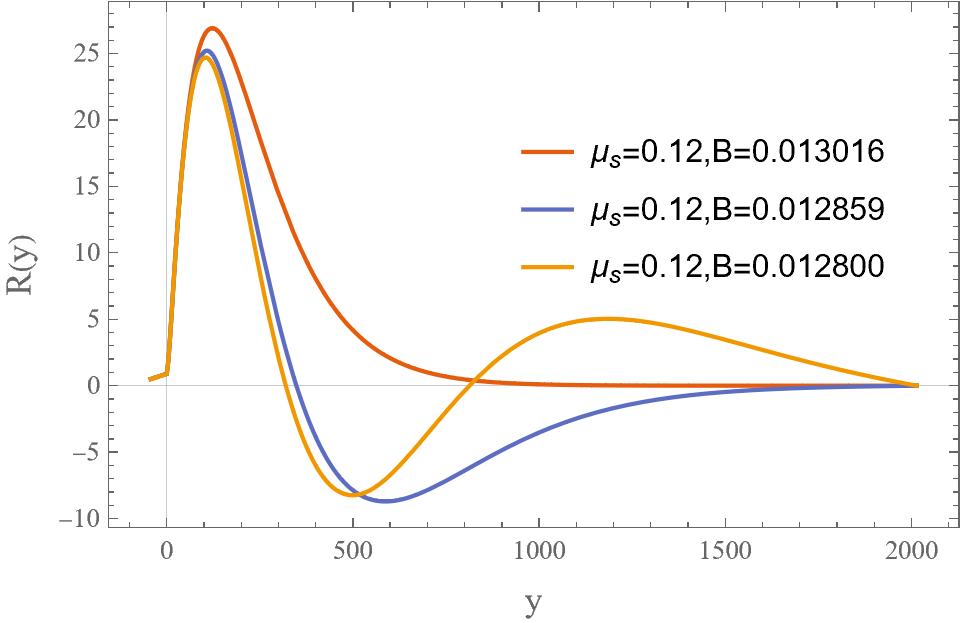} \hfill
    \includegraphics[width=0.45\textwidth]{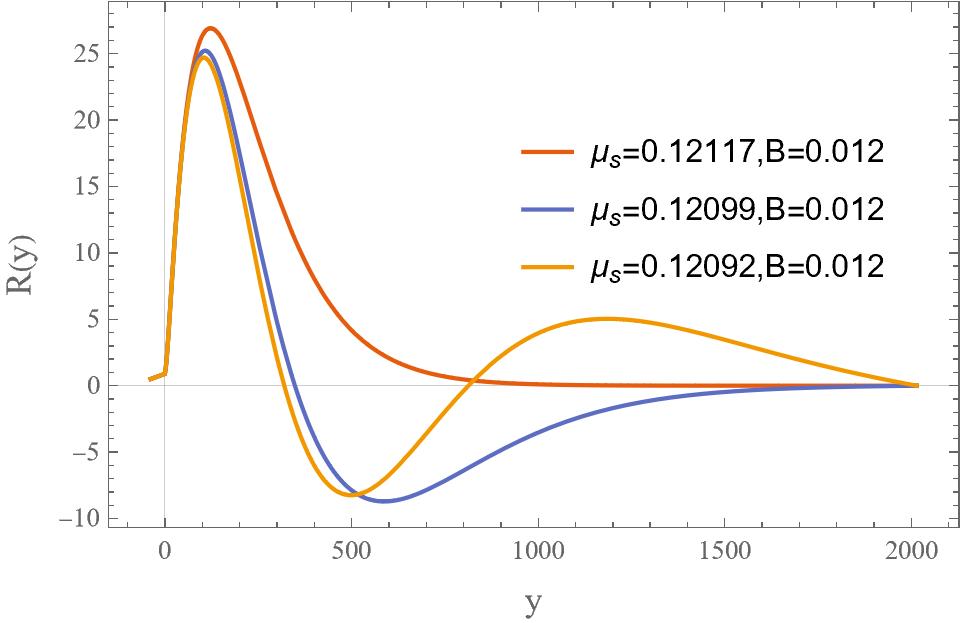}
    \caption{Comparison of different overtone modes ($n=0, 1, 2$) for the horizon-decaying scalar clouds at $\omega = \omega_c$. Left: Fixed scalar mass $\mu_s \approx 0.12$. Right: Fixed magnetic field $B = 0.012$.}
    \label{fig:compare_overtones_type2}
\end{figure}

The overtone modes ($n=1, 2$) for Type-II clouds are shown in Fig.~\ref{fig:compare_overtones_type2}. Similar to the Type-I branch, supporting higher radial nodes demands a delicate re-balancing of the potential well. The left panel shows that for a fixed $\mu_s$, slightly reducing the magnetic field $B$ allows the field to form additional nodes. The right panel demonstrates that fixing $B$ and slightly reducing $\mu_s$ achieves the same effect. This indicates the precise balance between the mass term and the magnetic confinement required to host higher-order horizon-decaying clouds.

\begin{figure}[htbp]
    \centering
    \includegraphics[width=0.45\textwidth]{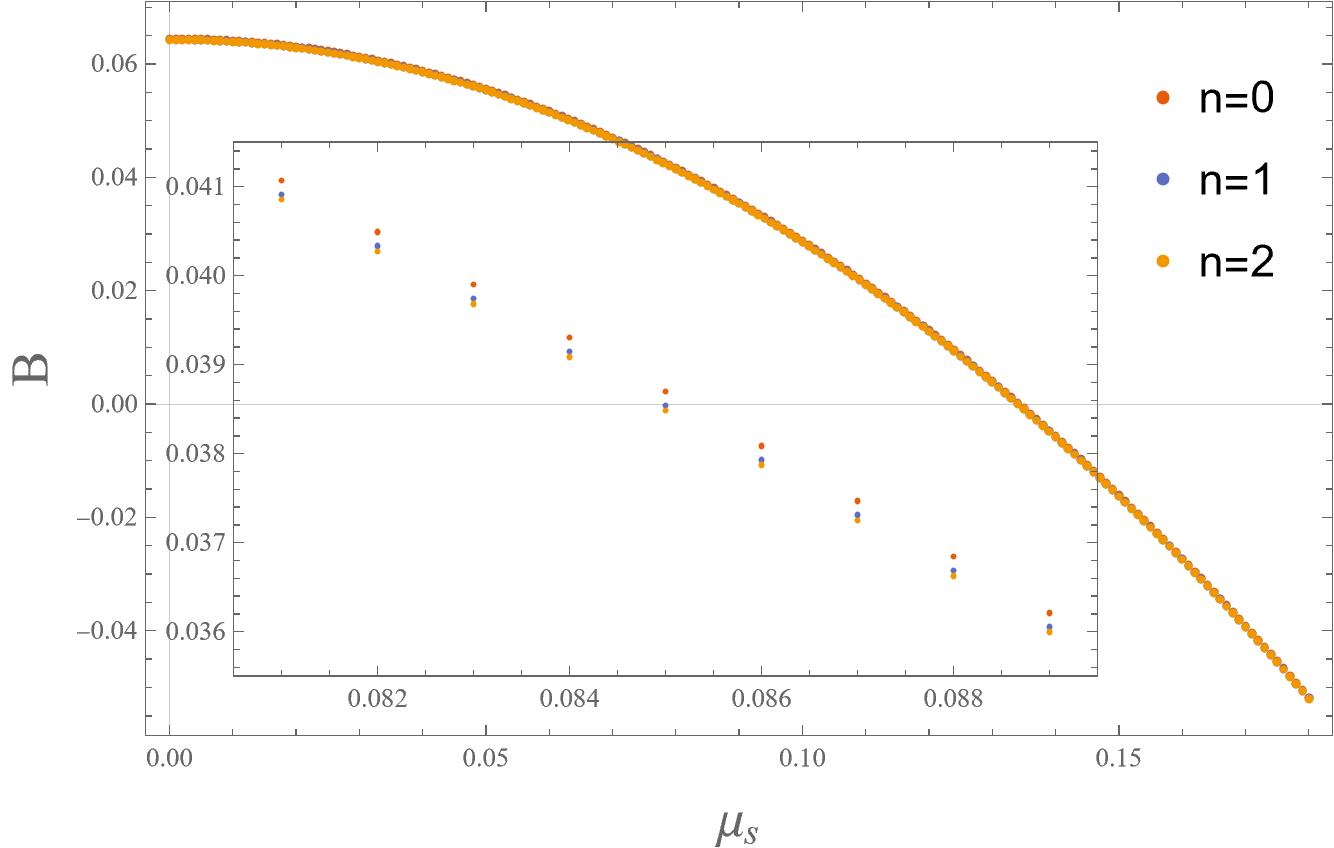} \hfill
    \includegraphics[width=0.45\textwidth]{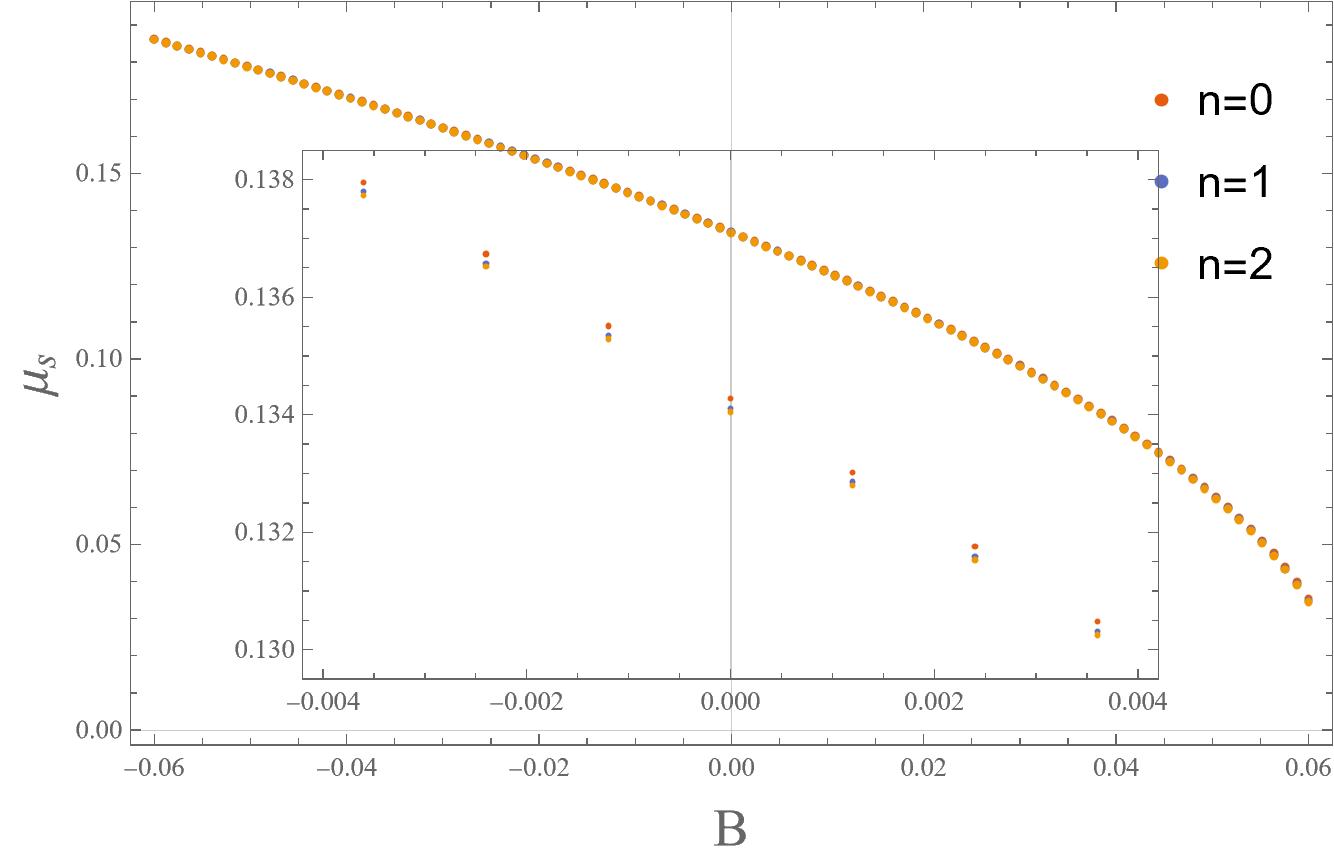} \\
    \vspace{0.2cm}
    \includegraphics[width=0.45\textwidth]{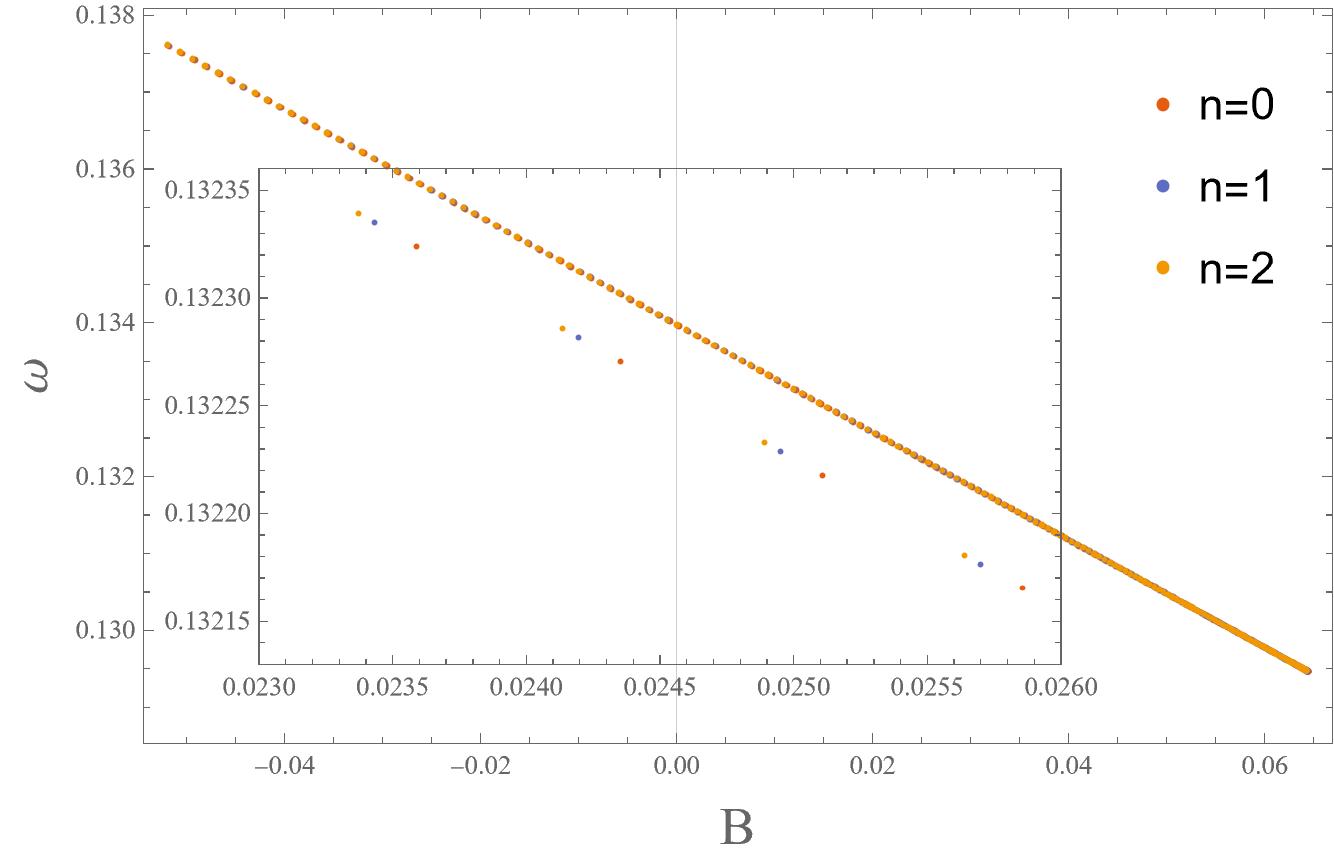} \hfill
    \includegraphics[width=0.45\textwidth]{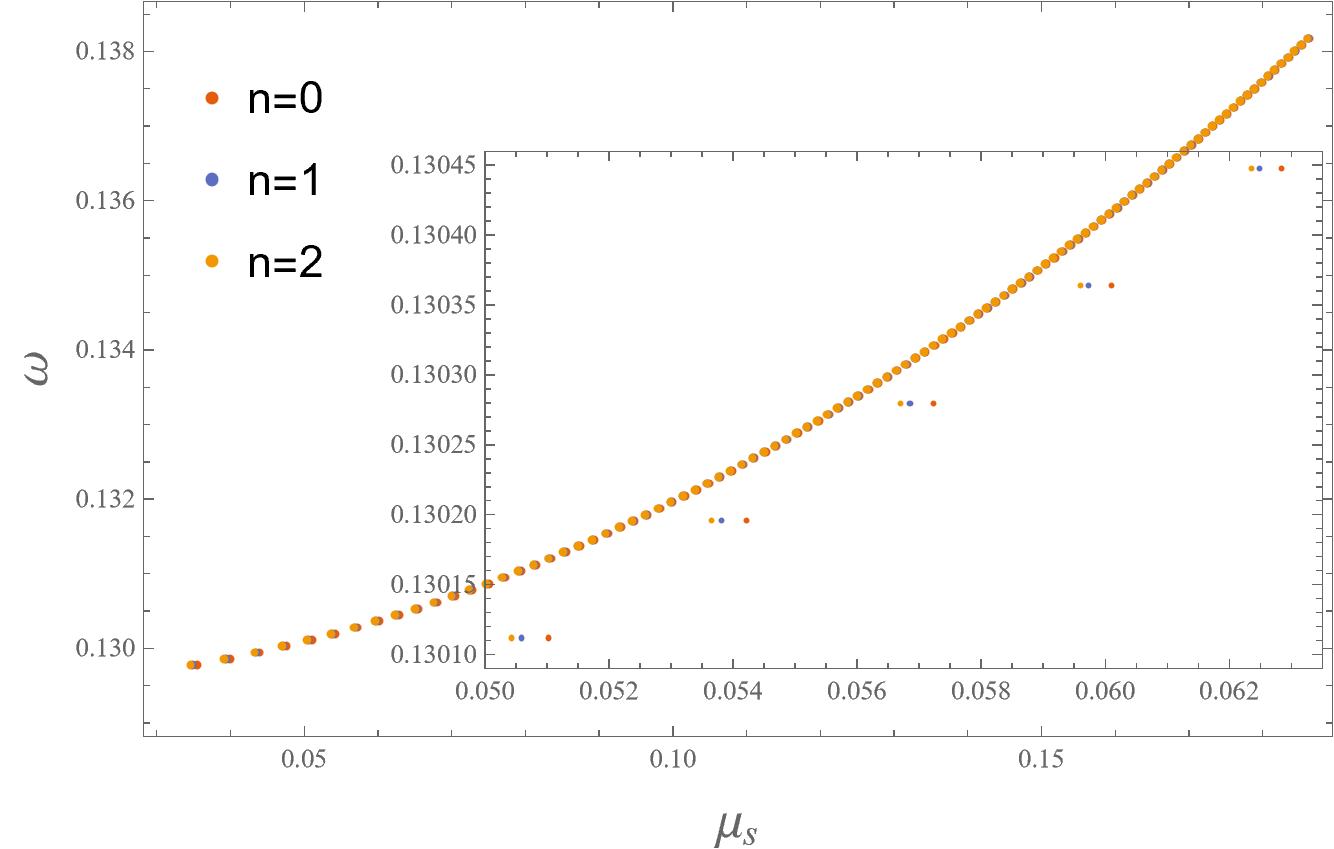}
    \caption{The permissible parameter space for Type-II horizon-decaying scalar clouds evaluated at $\omega = \omega_c$. The upper panels show the parameter bounds, and the lower panels show the frequency evolution. The insets resolve the discrete fine-structure splitting required to support higher nodal numbers.}
    \label{fig:evanescent_parameter_space}
\end{figure}

Finally, Figure~\ref{fig:evanescent_parameter_space} presents the parameter space and critical frequency evolution for Type-II clouds at $\omega=\omega_c$. The upper left panel displays the required $B$ values for different nodes when varying $\mu_s$, while the upper right panel shows the corresponding $\mu_s$ limits when varying $B$. A key physical constraint is observed here: to ensure $\mu_s > 0$ for these specific modes at $\omega=\omega_c$, the external magnetic field must be restricted to $B < 0.064482$. The lower panels show the evolution of $\omega_c$ with respect to $B$ (left) and $\mu_s$ (right). The insets resolve the discrete fine-structure splitting between different modes, confirming that the horizon-decaying branch is quantized in the same bound-state sense as ordinary scalar clouds while obeying a different horizon boundary condition.

\section{Discussion and Outlook}
\label{sec:conclusions}

In this work, we have identified a mechanism by which black-hole rotation and an external magnetic field jointly generate horizon-decaying scalar clouds. The central criterion is the appearance of a positive horizon gap in the near-horizon radial dispersion relation. The Kerr-Bertotti-Robinson geometry was used as a controlled separable setting in which this criterion can be realized and isolated analytically. By deriving the near-horizon boundary conditions and employing matched asymptotic expansions, we showed that the magnetic coupling modifies the rotational superradiant channel and creates a finite evanescent band in the bound-state spectrum.

Specifically, the fundamental manifestation of this rotation--magnetic quenching is that the propagating superradiant branch terminates \textit{before} the scalar field frequency reaches the kinematic synchronization threshold $\omega_c$. This premature termination creates a transition band ($\omega_1 < \omega < \omega_2$) in which the horizon wavenumber is imaginary. Consequently, the system supports two stationary configurations with different horizon physics:

\textbf{Type-I: Propagating reference clouds} \\
These clouds lie on the lower boundary of the quenching transition band ($\omega = \omega_1$). They are the zero-wavenumber edge of the propagating branch. Their near-horizon amplitude approaches a non-zero plateau, providing a useful reference against which the horizon-decaying solutions can be distinguished.

\textbf{Type-II: Horizon-decaying scalar clouds} \\
Inside the transition band ($\omega_1 < \omega < \omega_2$), the effective wavenumber at the horizon becomes purely imaginary ($k_h^2 < 0$). The scalar field therefore obeys an evanescent boundary condition and decays exponentially toward the horizon. This is the principal horizon-decaying branch found here. Its existence is tied to the simultaneous presence of rotation and magnetic coupling: rotation provides the superradiant channel, while the magnetic field reshapes the near-horizon dispersion relation so that the channel is converted into a zero-flux evanescent state. In the language of the general criterion, the Kerr-BR calculation supplies an explicit positive-gap realization, rather than merely an additional example of synchronized clouds. The small decay constant $\kappa_h$ gives these clouds a characteristic near-horizon profile that appears as a gradual outward growth in the numerical solutions.

The positive-gap criterion proposed here should be understood as a conditional near-horizon statement. It applies when the charged scalar equation can be separated, or reduced near the horizon after a regular field redefinition, to a radial equation of the form in Eq.~\eqref{eq:general_gap}. Establishing whether the same gap appears in fully nonseparable magnetized Kerr geometries, including higher-order magnetic backreaction and angular mode mixing, remains an open problem.

The weak-field Melvin--Kerr result \cite{siahaan2025melvinkerrclouds,siahaan2025melvinkerrsuperradiance} provides a useful external check on this formulation. It realizes the same two-root structure, although with a root center shifted relative to the synchronization frequency, and therefore supports the broader criterion without by itself proving the existence of a Type-II global cloud. The geometry-specific Kerr--BR calculation remains the explicit construction in which the evanescent near-horizon condition is matched to confinement at infinity.

This comparison also motivates extending the analysis beyond idealized uniform magnetic backgrounds. Nonuniform field geometries generated by accretion flows, jets, and plasma structures may produce the same local horizon root splitting, provided that a suitable near-horizon reduction remains valid. The existence of a local gap would then be a necessary diagnostic, while global confinement and regularity would still have to be established separately for each astrophysical geometry.

Our current analytical and numerical frameworks are based on the linear test-field approximation in the frequency domain. A natural next step is to test the time-domain evolution and non-linear stability of the Type-II branch. In particular, it would be important to determine whether the horizon-evanescent boundary condition remains robust under dynamical perturbations and backreaction.

Furthermore, our analytical mapping relies on the dimensionless weak-field condition $|B|M\ll1$. In geometrized units, the characteristic curvature-backreaction scale $B\sim M^{-1}$ corresponds to an extreme physical field of order $10^{19}(M_\odot/M)\,\mathrm{G}$ \cite{Galtsov:1978ag,zhang2024energy}. This gravitational weak-field criterion should be distinguished from the dynamical importance of a magnetic field for charged matter: the leading $O(qB)$ interaction can remain relevant while the $O(B^2)$ deformation of the metric is perturbatively small. Available astrophysical estimates illustrate this hierarchy. For example, horizon-scale polarimetric observations of M87* give a model-dependent radiating-plasma field of order $1$--$30\,\mathrm{G}$ \cite{EHT2021Magnetic}. Such a local plasma field is not identical to the uniform Kerr--BR background parameter, but it shows that astrophysically important fields can nevertheless be weak in the gravitational sense. Recent Kerr--BR analyses likewise organize geodesics and shadows as small-$B$ expansions \cite{Wang2026Shadows}, while the M87* jet-precession study gives the phenomenological bound $|B|M\lesssim0.0145$ \cite{Wang2026M87}. The value $BM=0.012$ used in several numerical examples therefore lies within the perturbative range, but should be regarded as a representative proof-of-principle choice rather than a direct measurement of the magnetic field of M87*. The same qualification applies to the wider parameter scans.

Exploring the deformation of the Type-II branch in ultra-strong magnetic fields (e.g., $B \sim 0.1/M$), where the $\mathcal{O}(B^2)$ spacetime backreaction and higher-order angular couplings cannot be perturbatively decoupled, remains an open theoretical challenge. Investigating whether these horizon-decaying clouds can seed non-linear hair or generate observational signatures would clarify the broader relevance of the rotation--magnetic-field mechanism. The main outcome is therefore a boundary-condition mechanism for stationary clouds, in which zero flux is achieved not by exact synchronization of a propagating mode but by converting the near-horizon channel into an evanescent one.
\appendix
\section{Technical Details: Field Equations}
\label{app:full_equations}

Under the weak-field assumption $|B|M \ll 1$ (written as $B \ll 1$ after the $M=1$ rescaling), and with the electromagnetic phase parameter $\xi = 0$, separating the variables $\Phi = \sum_{l,m} e^{i(m\phi - \omega t)} R(r) \, S(\theta)$ yields the leading-order weak-field radial and angular equations retained in our separable approximation. At generic $B$, the mass term breaks conformal covariance and, together with the non-aligned exact electromagnetic coupling, mixes $r$ and $\theta$. At the retained weak-field order, the leading $O(qB)$ terms separate and higher-order mixing is omitted. The equations below therefore describe an approximate, rather than exact, separable sector.

The radial equation is given by:
\begin{equation}
\begin{aligned}
& \frac{1}{R(r)} \frac{d}{dr}\left[ (a^2 - 2M r + r^2) R'(r) \right] \\
&\quad + \frac{1}{2} \mu_s^2 \left( -\frac{a^4}{a^2 + r(-2M + r)} + \frac{a^2 (2M - 3r) r}{a^2 + r(-2M + r)} + \frac{2 (2M - r) r^3}{a^2 + r(-2M + r)} + a^2 \right) \\
&\quad + \frac{1}{2} \omega^2 \left( \frac{a^4}{a^2 + r(-2M + r)} + \frac{2r^4}{a^2 + r(-2M + r)} + \frac{a^2 r (2M + 3r)}{a^2 + r(-2M + r)} - a^2 \right) \\
&\quad + \omega \left( -\frac{4a m M r}{a^2 + r(-2M + r)} + \frac{1}{2} \left( \frac{2a^5 Bq}{a^2 + r(-2M + r)}+ \frac{4a Bq r^3 (-M + r)}{a^2 + r(-2M + r)} \right. \right. \\
&\quad \left. \left. + \frac{2a^3 Bq r (-2M + 3r)}{a^2 + r(-2M + r)} - 2a^3 Bq \right) \right) \\
&\quad + \frac{1}{2} m^2 \frac{2a^2}{a^2 + r(-2M + r)} + \frac{1}{2} m \left( \frac{a^4 Bq}{a^2 + r(-2M + r)} + \frac{2Bq (2M - r) r^3}{a^2 + r(-2M + r)} \right. \\
&\quad \left. - \frac{a^2 Bq r (2M + 3r)}{a^2 + r(-2M + r)} + a^2 Bq \right) - K = 0.
\end{aligned}
\end{equation}

The corresponding angular equation takes the form of a generalized prolate spheroidal wave equation:
\begin{equation}
\begin{aligned}
& \frac{1}{S(\theta) \sin\theta} \frac{d}{d\theta} \left( \sin\theta S'(\theta) \right) - \frac{1}{2} \mu_s^2 a^2 \cdot 2 \cos^2\theta + \frac{1}{2} \omega^2 a^2 \cdot 2 \cos^2\theta\\
&\quad+ \frac{1}{2} \omega \cdot 2 a^3 B q \cdot 2 \cos^2\theta - \frac{1}{2} m^2 \cdot 2 \csc^2\theta - \frac{1}{2} m a^2 B q \cdot 2 \cos^2\theta - \frac{1}{2} m \cdot 2 a^2 B q \csc^2\theta + K=0.
\end{aligned}
\end{equation}

Introducing the coordinate $x=\cos\theta$ and defining the effective squared azimuthal parameter
\begin{equation}
\widetilde{m}^{\,2}\equiv m^2+a^2Bmq,
\label{eq:effective_azimuthal_parameter}
\end{equation}
the angular equation can be written as
\begin{equation}
\frac{d}{dx}\left[(1-x^2)\frac{dS}{dx}\right]
+\left[K-c^2x^2-\frac{\widetilde{m}^{\,2}}{1-x^2}\right]S=0,
\label{eq:generalized_spheroidal}
\end{equation}
where
\begin{equation}
c^2\equiv a^2\bigl(\mu_s^2-\omega^2\bigr)
+a^2Bq\bigl(m-2a\omega\bigr)
\label{eq:spheroidicity}
\end{equation}
is the effective spheroidicity parameter. In the weak-field regime, $\widetilde{m}^{\,2}=m^2+\mathcal{O}(B)$, so the magnetic contribution to the effective azimuthal parameter can be treated perturbatively. In the limit $B\to0$, the radial and angular equations recover their usual forms for a massive scalar field in Kerr spacetime; if in addition $c^2\to0$, the angular equation reduces to the associated Legendre equation.

When $|c^2|\ll1$, the spheroidal term $c^2 x^2$ can be treated as a perturbation. In this regime, using standard perturbation theory \cite{abramowitz1964handbook, zouros1979instabilities} and the separation-constant convention adopted in the radial equation, the generalized eigenvalue separation constant $K$ can be approximated to first order in $c^2$ as:
\begin{equation}
K \simeq l(l+1) + \frac{2c^2\left(m^2 + \frac{1}{2} - l(l+1)\right)}{(2l+3)(2l-1)} + a\bigl(2m\omega+a(\mu_s^2-\omega^2)\bigr).
\end{equation}

Furthermore, to analyze the physical boundary conditions and the properties of the bound states, it is useful to map the radial equation into a one-dimensional Schr\"{o}dinger-like form. By redefining the radial wave function as $Y(r) = \sqrt{r^2 + a^2} \, R(r)$, the radial equation can be recast as:
\begin{equation}
\frac{a^2 - 2M r + r^2}{r^2 + a^2} \cdot \frac{d}{dr} \left[ \frac{a^2 - 2M r + r^2}{r^2 + a^2} \cdot \frac{dY(r)}{dr} \right] - V_{\text{eff}}(r) \cdot Y(r) = 0,
\label{eq:yreq_explicit}
\end{equation}
where $V_{\text{eff}}(r)$ is the effective potential within the retained weak-field approximation. Collecting all the coupled terms, its full expression in this approximation is:
\begin{equation}
\begin{aligned}
V_{\text{eff}}(r) =& -\Biggl\{ -\frac{r(-2M+2r)(a^2-2Mr+r^2)}{(a^2+r^2)^{3/2}} + \frac{3r^2(a^2-2Mr+r^2)^2}{(a^2+r^2)^{5/2}} - \frac{(a^2-2Mr+r^2)^2}{(a^2+r^2)^{3/2}} \\
&\quad + \frac{a^2-2Mr+r^2}{\sqrt{a^2+r^2}} \Biggl[ -K + \frac{a^2 m^2}{a^2+r(-2M+r)} \\
&\quad + \frac{m}{2} \left( a^2 B q + \frac{a^4 B q}{a^2+r(-2M+r)} + \frac{2B q(2M-r)r^3}{a^2+r(-2M+r)} - \frac{a^2 B q r(2M+3r)}{a^2+r(-2M+r)} \right) \\
&\quad + \frac{\mu_s^2}{2} \left( a^2 - \frac{a^4}{a^2+r(-2M+r)} + \frac{a^2(2M-3r)r}{a^2+r(-2M+r)} + \frac{2(2M-r)r^3}{a^2+r(-2M+r)} \right) \\
&\quad + \omega \Biggl( -\frac{4a m M r}{a^2+r(-2M+r)} + \frac{1}{2}\biggl( -2a^3 B q + \frac{2a^5 B q}{a^2+r(-2M+r)} \\
&\quad + \frac{4a B q r^3(-M+r)}{a^2+r(-2M+r)} + \frac{2a^3 B q r(-2M+3r)}{a^2+r(-2M+r)} \biggr) \Biggr) \\
&\quad + \frac{\omega^2}{2} \left( -a^2 + \frac{a^4}{a^2+r(-2M+r)} + \frac{2r^4}{a^2+r(-2M+r)} + \frac{a^2 r(2M+3r)}{a^2+r(-2M+r)} \right) \Biggr] \Biggr\} \\
&\quad \times \frac{1}{(a^2+r^2)^{3/2}}.
\end{aligned}
\label{eq:veff_explicit}
\end{equation}
\section{Technical Details: Bound-State Constraints}
\label{app:boundary_conditions}

The physical horizon boundary condition is discussed in Sec.~\ref{sec:effective_potential}. Here we collect the remaining bound-state constraints at spatial infinity, which determine when the radial solution is localized rather than radiative.

To ensure the scalar field forms a bound state rather than radiating to infinity, $k_f^2 > 0$ must be strictly maintained. This imposes the following parameter space constraints on the scalar mass $\mu_s$ and azimuthal number $m \in \mathbb{Z}$:

\noindent \textbf{If $-1 < q < 0$:}
\begin{equation}
\begin{cases}
-1 < B < 0 \text{ and } a > 0 \text{ and }
\begin{cases}
m \leq 0 \text{ and } M > a \text{ and } \mu_s > \sqrt{-B m q} \\
m \geq 1 \text{ and } M > a \text{ and } \mu_s \geq 0
\end{cases} \\
B = 0 \text{ and } a > 0 \text{ and } M > a \text{ and } \mu_s > 0 \\
0 < B < 1 \text{ and } a > 0 
\text{ and }
\begin{cases}
m \leq -a^2 B q \text{ and } M > a \text{ and } \mu_s \geq 0 \\
m > -a^2 B q \text{ and } M > a \text{ and } \mu_s \geq \sqrt{-a^2 B^2 q^2 - B m q}
\end{cases}
\end{cases}
\end{equation}

\noindent \textbf{If $q = 0$:}
\begin{equation}
-1 < B < 1 \text{ and } a > 0 \text{ and } M > a \text{ and } \mu_s > 0
\end{equation}

\noindent \textbf{If $0 < q < 1$:}
\begin{equation}
\begin{cases}
-1 < B < 0 \text{ and } a > 0 \text{ and }
\begin{cases}
m \leq -a^2 B q \text{ and } M > a \text{ and } 
\mu_s \geq 0 \\
m > -a^2 B q \text{ and } M > a \text{ and } \mu_s \geq \sqrt{-a^2 B^2 q^2 - B m q}
\end{cases} \\
B = 0 \text{ and } a > 0 \text{ and } M > a \text{ and } \mu_s > 0 \\
0 < B < 1 \text{ and } a > 0 \text{ and }
\begin{cases}
m \leq 0 \text{ and } M > a \text{ and } \mu_s > \sqrt{-B m q} \\
m \geq 1 \text{ and } M > a \text{ and } \mu_s \geq 0
\end{cases}
\end{cases}
\end{equation}

\section{Technical Details: Matched Asymptotic Expansion}
\label{app:matching}

We determine the complex frequency $\omega$ by a matched asymptotic expansion \cite{detweiler1980klein,furuhashi2004instability}, distinct from the weak-$B$ truncation. In the weak-binding regime $|k|M\ll1$, the near solution applies for $r-r_+\ll|k|^{-1}$, the far solution for $r-r_+\gg M$, and they overlap in $M\ll r-r_+\ll|k|^{-1}$. Since $r_+-r_-\leq2M$, the weak-binding condition also implies $|k|(r_+-r_-)\ll1$.

For $r-r_+ \gg M$, the radial equation is approximately:
\begin{equation}
\frac{d^2}{dr^2}\,(rR) + \left[k^2 - \frac{2i\nu k}{r} - \frac{l(l+1)+\epsilon^2}{r^2}\right] rR = 0, \label{eq:outer_approx}
\end{equation}
with definitions
\begin{align}
k &\equiv \sqrt{-B m q - \mu_s^2 + 2 a B q \omega + \omega^2}, \\
\nu &\equiv -i \frac{B m M q + M \mu_s^2 - 3 a B M q \omega - 2 M \omega^2}{k},
\qquad x=2kr .
\end{align}
Here $\epsilon^2$ denotes the subleading correction to the
far-region centrifugal coefficient. It is retained as a bookkeeping term in
the Gamma-function matching and omitted from the leading spectrum.
The solution regular at infinity is
\begin{equation}
R(x)= e^{i x/2} x^{l} \, {}_2 U\left(1 + l + \epsilon^2 - \nu,\ 2(1 + l + \epsilon^2),\ -i x\right),
\end{equation}
where ${}_2 U$ is the confluent hypergeometric function of the second kind \cite{abramowitz1964handbook}. Its asymptotic behavior for $|kr| \ll 1$ is:
\begin{equation}
R \sim \frac{2^l (kr)^l \Gamma\!\left(-1-2l-2\epsilon^2\right)}{\Gamma\!\left(-l-\epsilon^2-\nu\right)} - \frac{2^{-1-l} (kr)^{-1-l} \Gamma\!\left(1+2l+2\epsilon^2\right)}{e^{2il\pi}\,\Gamma\!\left(1+l+\epsilon^2-\nu\right)}. \label{asymptotic_near}
\end{equation}

Conversely, in the near region $r-r_+\ll |k|^{-1}$, transforming to $z=(r-r_+)/(r_+ -r_-)$ simplifies the equation to:
\begin{equation}
z(z+1)\frac{d}{dz}\left[z(z+1)\frac{dR}{dz}\right] + \bigl[\alpha^2 - l(l+1)z(z+1)\bigr]R = 0,
\end{equation}
where the near-horizon coupling parameter $\alpha^2$ is explicitly given by:
\begin{equation}
\begin{aligned}
\alpha^2 \equiv & \frac{1}{4(a^2-M^2)} \Bigl[ -a^4 B m q + a^3 B M q \omega (6M - r_- + r_+) \\
& + a^2 \bigl(2 B m M q r_+ - m^2 + 4 M^2 \omega^2 \bigr) + 4 a M r_+ \omega (m - 2 B M^2 q) - 8 M^3 r_+ \omega^2 \Bigr].
\end{aligned}
\end{equation}
When $|B|M \ll 1$ and outside the quenching band ($\omega<\omega_1$ or $\omega>\omega_2$), $\alpha$ is real and can be approximated as $\alpha \simeq \alpha_0+\frac{\alpha_B^2}{2\alpha_0}$, where
\begin{equation}
\alpha_0 = \frac{a m-2 M r_+ \omega }{r_+-r_-}, \quad \alpha_B = \frac{1}{2} \sqrt{-\frac{a B q \left(a^3 m+a^2 M \omega  (-6 M+r_- -r_+)-2 a m M r_+ +8 M^3 r_+ \omega \right)}{a^2-M^2}}.
\end{equation}

The solution with the $z^{i\alpha}$ near-horizon behavior is
\begin{equation}
R(z)=\left(\frac{z}{1+z}\right)^{i\alpha}
F(-l,l+1;1+2i\alpha;-z).
\end{equation}

Here $F$ is the Gauss hypergeometric function
\cite{abramowitz1964handbook}. Its large-$z$ asymptotic form ($z\gg1$) is
\begin{equation}
R \sim \frac{\Gamma(1 + 2i\alpha)\Gamma(2l + 1)}{\Gamma(l + 1)\Gamma(l + 1 + 2i\alpha)} \left( \frac{r}{r_+ - r_-} \right)^l + \frac{\Gamma(1 + 2i\alpha)\Gamma(-2l - 1)}{\Gamma(-l)\Gamma(-l + 2i\alpha)} \left( \frac{r}{r_+ - r_-} \right)^{-l-1}. \label{asymptotic_far}  
\end{equation}

In the overlap region $M\ll r-r_+\ll|k|^{-1}$, matching the leading-order terms of Eq.~\eqref{asymptotic_near} and \eqref{asymptotic_far} yields the resonance condition \cite{detweiler1980klein, furuhashi2004instability}:
\begin{equation}
    \frac{\Gamma(-l - \nu - \epsilon^2) \Gamma(2l + 2)}{\Gamma(l - \nu + 1 + \epsilon^2) \Gamma(-2l - 2\epsilon^2)} = -2\alpha \left[ 2k(r_+ - r_-) \right]^{2l+1} \prod_{j=1}^l (j^2 + 4\alpha^2) \frac{\Gamma(l+1) \Gamma(-2l-1)}{\Gamma(2l+1)\Gamma(-l)}. \label{match}
\end{equation}
At fixed $\alpha$, the right-hand side is parametrically of order $[|k|(r_+-r_-)]^{2l+1}$. Setting it to zero gives the leading pole condition. Defining $\nu^{(0)}$ by $l-\nu^{(0)}+1+\epsilon^2=-n$, with $n=0,1,2,\ldots$, we obtain
\begin{equation}
    \nu^{(0)} = -\frac{i \left( B m M q + M \mu_s^2 - 3 a B M q \omega^{(0)} - 2 M {\omega^{(0)}}^2 \right)}{\sqrt{-B m q - \mu_s^2 + 2 a B q \omega^{(0)} + {\omega^{(0)} }^2}}\simeq l+n+1
\end{equation}
\begin{equation}
    \omega^{(0)}=\omega_f - \frac{M^2 (\omega_f^2-(a^2 B^2 q^2-a B q \sqrt{a^2 B^2 q^2+B m q+\mu_s^2}))^2}{2 (l+n+1)^2 \sqrt{a^2 B^2 q^2+B m q+{\mu_s}^2}},
\end{equation}
\begin{equation}    
    k^{(0)}=i \sqrt{\frac{M^2 \sqrt{a^2 B^2 q^2+B m q+{\mu_s}^2 } {\omega_f}^3}{(l+n+1)^2}},
\end{equation}
Restoring the right-hand side, we expand the quantization
condition as $\nu=\nu^{(0)}+\delta\nu$ and
$\omega=\omega^{(0)}+\delta\omega$. At fixed $\alpha^{(0)}$, the first
correction satisfies
$\delta\nu=O([|k^{(0)}|(r_+-r_-)]^{2l+1})$ and is
\begin{equation}
    \delta \nu = 2\alpha^{(0)} \left[ 2k^{(0)}(r_+ - r_-) \right]^{2l+1} \frac{(2l+1+n)!}{n!} \left[ \frac{l!}{(2l)!(2l+1)!} \right]^2 \prod_{j=1}^l \left( j^2 + 4{\alpha^{(0)}}^2 \right). \label{deltanu2}
\end{equation}
The frequency shift follows from
$\delta\nu=(\partial\nu/\partial\omega)_{\omega^{(0)}}\delta\omega$. Thus
$\omega=\omega^{(0)}+\delta\omega$, with leading real part
$\sigma=\omega^{(0)}$ and growth or decay rate
$\gamma=\operatorname{Im}(\delta\omega)$.

The crucial step occurs when investigating the exponentially decaying regime where ${\alpha^{(0)}}^2 < 0$. Defining $\beta^{(0)}\equiv\sqrt{-{\alpha^{(0)}}^2}>0$, the physical branch compatible with $R\sim z^{i\alpha}$ is
\[
\alpha^{(0)}=-i\beta^{(0)}.
\]

The leading-order asymptotic wavenumber is purely imaginary,
$k^{(0)}=i|k^{(0)}|$. Consequently, the factor involving $k^{(0)}$ yields an imaginary phase: $[k^{(0)}]^{2l+1} = i^{2l+1} |k^{(0)}|^{2l+1} = i (-1)^l |k^{(0)}|^{2l+1}$.

When the system enters the exponentially decaying regime, the linear factor $\alpha^{(0)}$ contributes an additional factor of $-i$. The product of these two imaginary phases yields:
\begin{equation}
    \alpha^{(0)} \left[ k^{(0)} \right]^{2l+1} \propto (-i\beta^{(0)}) \left( i (-1)^l |k^{(0)}|^{2l+1} \right) = (-1)^l\beta^{(0)}|k^{(0)}|^{2l+1}.
\end{equation}

Because $(-i)\times i=1$, the imaginary units exactly cancel out. Since the product term $\prod_{j=1}^l (j^2 + 4{\alpha^{(0)}}^2) = \prod_{j=1}^l (j^2 - 4{\beta^{(0)}}^2)$ remains strictly real, and the background derivative $\frac{\partial\nu}{\partial\omega}\bigg|_{(\omega=\omega^{(0)})}$ is also a real quantity, the frequency shift $\delta\omega$ is forced to be purely real. Therefore, $\gamma = \text{Im}(\delta\omega) = 0$.

\section{Technical Details: Numerical Calculation Strategy}
\label{app:numerical}

To numerically integrate the radial equation \eqref{eq:yreq}, the radial function $Y(r)$ is expanded in a Frobenius series near the event horizon to obtain the initial conditions:
\begin{equation}
    Y(r) = (r-r_+)^{s_H} \sum_{j=0}^{N} Y_j (r-r_+)^j, \qquad r \rightarrow r_+, \label{HorizonExpansion}
\end{equation}
where the characteristic exponent is $s_H=i\alpha$. On the Type-II branch, $\alpha=-i\beta$ and hence $s_H=\beta>0$, which selects the solution that vanishes at the event horizon. By inserting this expansion into the radial equation, the higher-order coefficients $Y_j$ ($j \geq 1$) can be solved recursively in terms of $Y_0$. Since the differential equation is linear, we fix $Y_0 = 1$ without loss of generality.

In practical numerical calculations, we truncate the infinite radial domain to $r \in [r_{+n}, r_{\text{INF}}]$, where $r_{+n} = r_+(1+\epsilon)$ is a point very close to the horizon with a small parameter $\epsilon$, and $r_{\text{INF}}$ is a sufficiently large cutoff. We use the near-horizon expansion evaluated at $r_{+n}$ to provide the initial values $Y(r_{+n})$ and $Y'(r_{+n})$. From this inner boundary, we perform a \textit{direct numerical integration} outward to $r_{\text{INF}}$.

The two kinds of scalar clouds are determined by a set of five parameters $\{a, q, \omega, \mu_s, B\}$. When the black hole spin $a$, scalar charge $q$, and field frequency $\omega$ are fixed, only specific discrete values of the remaining parameters $\{\mu_s, B\}$ allow for a bounded solution where $Y(r \rightarrow \infty) = 0$ to be achieved. Specifically, for fixed $\{a, q, \omega\}$, there exist two infinite countable sets of coupling constants, $\{\mu_s(a, q, \omega,B;n)\}_{n=0}^\infty$ and $\{B(a, q, \omega,\mu_s;n)\}_{n=0}^\infty$, which can support the bounded scalar clouds. The integer $n$ labels the number of radial nodes of the solution.

By employing a shooting method \cite{press2007numerical} over these parameter spaces, we iteratively adjust the tuning parameter (either $\mu_s$ or $B$) until the integrated wavefunction satisfies the asymptotic bound-state boundary condition, exponentially decaying to zero at the outer boundary ($Y(r_{\text{INF}}) \to 0$). For our computations, we typically set $N=3$, $\epsilon = 10^{-9}$, and $r_{\text{INF}} = 1500 r_+$.

\begin{acknowledgments}
	This work is supported by the National Natural Science Foundation of China (NNSFC) under Grant No 12075207.
\end{acknowledgments}

\bibliographystyle{utphys}
\bibliography{ref}

@article{Penrose:1969pc,
    author = "Penrose, R.",
    title = "{Gravitational collapse: The role of general relativity}",
    doi = "10.1023/A:1016578408204",
    journal = "Riv. Nuovo Cim.",
    volume = "1",
    pages = "252--276",
    year = "1969"
}

@article{christodoulou1970reversible,
  title={Reversible and irreversible transformations in black-hole physics},
  author={Christodoulou, Demetrios},
  journal={Physical Review Letters},
  volume={25},
  number={25},
  pages={1596--1597},
  year={1970},
  publisher={APS},
  doi={10.1103/PhysRevLett.25.1596}
}

@article{ruffini1971introducing,
  title={Introducing the black hole},
  author={Ruffini, Remo and Wheeler, John A},
  journal={Physics Today},
  volume={24},
  number={1},
  pages={30--41},
  year={1971},
  doi={10.1063/1.3022513}
}

@article{Zeldovich:1971ffh,
    author = "Zeldovich, Yakov Borisovich",
    title = "{Generation of Waves by a Rotating Body}",
    journal = "Soviet Journal of Experimental and Theoretical Physics Letters",
    volume = "14",
    pages = "180--181",
    year = "1971"
}

@article{Zeldovich:1972zqp,
    author = "Zeldovich, Yakov Borisovich",
    title = "{Amplification of Cylindrical Electromagnetic Waves Reflected from a Rotating Body}",
    journal = "Soviet Journal of Experimental and Theoretical Physics",
    volume = "35",
    number = "6",
    pages = "1085--1087",
    year = "1972"
}

@article{misner1972interpretation,
  title={Interpretation of gravitational-wave observations},
  author={Misner, Charles W},
  journal={Physical Review Letters},
  volume={28},
  number={15},
  pages={994--997},
  year={1972},
  publisher={APS},
  doi={10.1103/PhysRevLett.28.994}
}

@article{Starobinskii:1973vzb,
    author = "Starobinskii, A. A.",
    title = "{Amplification of waves during reflection from a rotating ''black hole''}",
    journal = "Sov. Phys. JETP",
    volume = "37",
    number = "1",
    pages = "28--32",
    year = "1973"
}

@article{bekenstein1973extraction,
  title={Extraction of energy and charge from a black hole},
  author={Bekenstein, Jacob D},
  journal={Physical Review D},
  volume={7},
  number={4},
  pages={949--953},
  year={1973},
  publisher={APS},
  doi={10.1103/PhysRevD.7.949}
}

@article{teukolsky1973perturbations,
  title={Perturbations of a rotating black hole. I. Fundamental equations for gravitational, electromagnetic, and neutrino-field perturbations},
  author={Teukolsky, Saul A},
  journal={The Astrophysical Journal},
  volume={185},
  pages={635--648},
  year={1973},
  doi={10.1086/152444}
}

@article{press1972floating,
  title={Floating orbits, superradiant scattering and the black-hole bomb},
  author={Press, William H and Teukolsky, Saul A},
  journal={Nature},
  volume={238},
  number={5361},
  pages={211--212},
  year={1972},
  publisher={Nature Publishing Group},
  doi={10.1038/238211a0}
}

@article{damour1976black,
  title={On quantum resonances in stationary geometries},
  author={Damour, Thibault and Deruelle, Nathalie and Ruffini, Remo},
  journal={Lett. Nuovo Cim.},
  volume={15},
  number={8},
  pages={257--262},
  year={1976},
  publisher={Springer},
  doi={10.1007/BF02725534}
}

@article{zouros1979instabilities,
  title={Instabilities of massive scalar perturbations of a rotating black hole},
  author={Zouros, T. J. M. and Eardley, Douglas M.},
  journal={Annals of Physics},
  volume={118},
  number={1},
  pages={139--155},
  year={1979},
  publisher={Elsevier},
  doi={10.1016/0003-4916(79)90237-9}
}

@article{detweiler1980klein,
  title={Klein-Gordon equation and rotating black holes},
  author={Detweiler, Steven},
  journal={Physical Review D},
  volume={22},
  number={10},
  pages={2323--2326},
  year={1980},
  publisher={APS},
  doi={10.1103/PhysRevD.22.2323}
}

@article{furuhashi2004instability,
  title={Instability of massive scalar fields in Kerr-Newman spacetime},
  author={Furuhashi, Hironobu and Nambu, Yasusada},
  journal={Progress of Theoretical Physics},
  volume={112},
  number={6},
  pages={983--995},
  year={2004},
  publisher={Oxford University Press},
  doi={10.1143/PTP.112.983}
}

@article{cardoso2004black,
  title={Black-hole bomb and superradiant instabilities},
  author={Cardoso, Vitor and Dias, Oscar J. C. and Lemos, Jos{\'e} P. S. and Yoshida, Shijun},
  journal={Physical Review D},
  volume={70},
  number={4},
  pages={044039},
  year={2004},
  publisher={APS},
  doi={10.1103/PhysRevD.70.044039}
}

@article{dolan2007instability,
  title={Instability of the massive Klein-Gordon field on the Kerr spacetime},
  author={Dolan, Sam R.},
  journal={Physical Review D},
  volume={76},
  number={8},
  pages={084001},
  year={2007},
  publisher={APS},
  doi={10.1103/PhysRevD.76.084001}
}

@article{dolan2013superradiant,
  title={Superradiant instabilities of rotating black holes in the time domain},
  author={Dolan, Sam R.},
  journal={Physical Review D},
  volume={87},
  number={12},
  pages={124026},
  year={2013},
  publisher={APS},
  doi={10.1103/PhysRevD.87.124026}
}

@article{witek2012superradiant,
  title={Superradiant instabilities in astrophysical systems},
  author={Witek, Helvi and Cardoso, Vitor and Ishibashi, Akihiro and Sperhake, Ulrich},
  journal={Physical Review D},
  volume={87},
  number={4},
  pages={043513},
  year={2013},
  publisher={APS},
  doi={10.1103/PhysRevD.87.043513}
}

@article{hod2012stationary,
  title={Stationary scalar clouds around rotating black holes},
  author={Hod, Shahar},
  journal={Physical Review D},
  volume={86},
  number={10},
  pages={104026},
  year={2012},
  publisher={APS},
  doi={10.1103/PhysRevD.86.104026}
}

@article{hod2014kerr,
  title={Kerr-Newman black holes with stationary charged scalar clouds},
  author={Hod, Shahar},
  journal={Physical Review D},
  volume={90},
  number={2},
  pages={024051},
  year={2014},
  publisher={APS},
  doi={10.1103/PhysRevD.90.024051}
}

@article{herdeiro2014kerr,
  title={Kerr black holes with scalar hair},
  author={Herdeiro, Carlos A. R. and Radu, Eugen},
  journal={Physical Review Letters},
  volume={112},
  number={22},
  pages={221101},
  year={2014},
  publisher={APS},
  doi={10.1103/PhysRevLett.112.221101}
}

@article{herdeiro2015construction,
  title={Construction and physical properties of Kerr black holes with scalar hair},
  author={Herdeiro, Carlos A. R. and Radu, Eugen},
  journal={Classical and Quantum Gravity},
  volume={32},
  number={14},
  pages={144001},
  year={2015},
  publisher={IOP Publishing},
  doi={10.1088/0264-9381/32/14/144001}
}

@article{Herdeiro:2015waa,
    author = "Herdeiro, Carlos A. R. and Radu, Eugen",
    title = "{Asymptotically flat black holes with scalar hair: a review}",
    eprint = "1504.08209",
    archivePrefix = "arXiv",
    primaryClass = "gr-qc",
    doi = "10.1142/S0218271815420146",
    journal = "Int. J. Mod. Phys. D",
    volume = "24",
    number = "09",
    pages = "1542014",
    year = "2015"
}

@article{benone2014kerr,
  title={Kerr-Newman scalar clouds},
  author={Benone, Carolina L. and Crispino, Lu{\'\i}s C. B. and Herdeiro, Carlos A. R. and Radu, Eugen},
  journal={Physical Review D},
  volume={90},
  number={10},
  pages={104024},
  year={2014},
  publisher={APS},
  doi={10.1103/PhysRevD.90.104024}
}

@article{Huang:2017whw,
    author = "Huang, Yang and Liu, Dao-Jun and Zhai, Xiang-Hua and Li, Xin-Zhou",
    title = "{Scalar clouds around Kerr{\textendash}Sen black holes}",
    eprint = "1706.04441",
    archivePrefix = "arXiv",
    primaryClass = "gr-qc",
    doi = "10.1088/1361-6382/aa7964",
    journal = "Class. Quant. Grav.",
    volume = "34",
    number = "15",
    pages = "155002",
    year = "2017"
}

@article{pani2012black,
  title={Black-hole bomb and photon-mass bounds},
  author={Pani, Paolo and Cardoso, Vitor and Gualtieri, Leonardo and Berti, Emanuele and Ishibashi, Akihiro},
  journal={Physical Review Letters},
  volume={109},
  number={13},
  pages={131102},
  year={2012},
  publisher={APS},
  doi={10.1103/PhysRevLett.109.131102}
}

@article{PhysRevLett.119.041101,
  title = {Superradiant Instability and Backreaction of Massive Vector Fields around Kerr Black Holes},
  author = {East, William E. and Pretorius, Frans},
  journal = {Phys. Rev. Lett.},
  volume = {119},
  issue = {4},
  pages = {041101},
  numpages = {5},
  year = {2017},
  month = {Jul},
  publisher = {American Physical Society},
  doi = {10.1103/PhysRevLett.119.041101},
  url = {https://link.aps.org/doi/10.1103/PhysRevLett.119.041101}
}

@article{herdeiro2016kerr,
  title={Kerr black holes with Proca hair},
  author={Herdeiro, Carlos and Radu, Eugen and R{\'u}narsson, Helgi},
  journal={Classical and Quantum Gravity},
  volume={33},
  number={15},
  pages={154001},
  year={2016},
  publisher={IOP Publishing},
  doi={10.1088/0264-9381/33/15/154001}
}

@article{hod2021stationary,
  title={Stationary scalar clouds supported by rapidly-rotating acoustic black holes in a photon-fluid model},
  author={Hod, Shahar},
  journal={Physical Review D},
  volume={103},
  number={8},
  pages={084003},
  year={2021},
  publisher={APS},
  doi={10.1103/PhysRevD.103.084003}
}

@article{ciszak2021acoustic,
  title={Acoustic black-hole bombs and scalar clouds in a photon-fluid model},
  author={Ciszak, M. and Marino, F.},
  journal={Physical Review D},
  volume={103},
  number={4},
  pages={045004},
  year={2021},
  publisher={APS},
  doi={10.1103/PhysRevD.103.045004}
}

@book{Brito:2015oca,
    author = "Brito, Richard and Cardoso, Vitor and Pani, Paolo",
    title = "{Superradiance}: {New Frontiers in Black Hole
Physics}",
    eprint = "1501.06570",
    archivePrefix = "arXiv",
    primaryClass = "gr-qc",
    doi = "10.1007/978-3-319-19000-6",
    series = "Lecture Notes in Physics",
    volume = "906",
    pages = "1--237",
    publisher = "Springer",
    year = "2015"
}

@article{blandford1977electromagnetic,
  title={Electromagnetic extraction of energy from Kerr black holes},
  author={Blandford, Roger D. and Znajek, Roman L.},
  journal={Monthly Notices of the Royal Astronomical Society},
  volume={179},
  number={3},
  pages={433--456},
  year={1977},
  publisher={Oxford University Press},
  doi={10.1093/mnras/179.3.433}
}

@article{wald1974black,
  title={Black hole in a uniform magnetic field},
  author={Wald, Robert M.},
  journal={Physical Review D},
  volume={10},
  number={6},
  pages={1680--1685},
  year={1974},
  publisher={APS},
  doi={10.1103/PhysRevD.10.1680}
}

@article{ernst1976black,
  title={Black holes in a magnetic universe},
  author={Ernst, Frederick J.},
  journal={Journal of Mathematical Physics},
  volume={17},
  number={1},
  pages={54--56},
  year={1976},
  publisher={American Institute of Physics},
  doi={10.1063/1.522781}
}

@article{ernstwild1976kerr,
  title={Kerr black holes in a magnetic universe},
  author={Ernst, Frederick J. and Wild, Walter J.},
  journal={Journal of Mathematical Physics},
  volume={17},
  number={2},
  pages={182},
  year={1976},
  publisher={American Institute of Physics},
  doi={10.1063/1.522875}
}

@article{melvin1964pure,
  title={Pure magnetic and magic electric geons},
  author={Melvin, M. A.},
  journal={Physics Letters},
  volume={8},
  number={1},
  pages={65--68},
  year={1964},
  publisher={Elsevier},
  doi={10.1016/0031-9163(64)90801-7}
}

@article{gibbons2013ergoregions,
  title={Ergoregions in magnetised black hole spacetimes},
  author={Gibbons, Gary W. and Mujtaba, Ali H. and Pope, Christopher N.},
  journal={Classical and Quantum Gravity},
  volume={30},
  number={12},
  pages={125008},
  year={2013},
  publisher={IOP Publishing},
  doi={10.1088/0264-9381/30/12/125008}
}

@article{booth2015insights,
  title={Insights from Melvin--Kerr--Newman spacetimes},
  author={Booth, Ivan and Hunt, Matthew and Palomo-Lozano, Alberto and Kunduri, Hari K.},
  journal={Classical and Quantum Gravity},
  volume={32},
  number={23},
  pages={235025},
  year={2015},
  publisher={IOP Publishing},
  doi={10.1088/0264-9381/32/23/235025}
}

@article{Konoplya:2008hj,
    author = "Konoplya, R. A.",
    title = "{Magnetic field creates strong superradiant instability}",
    eprint = "0801.0846",
    archivePrefix = "arXiv",
    primaryClass = "hep-th",
    doi = "10.1016/j.physletb.2008.11.059",
    journal = "Phys. Lett. B",
    volume = "666",
    pages = "283--287",
    year = "2008"
}

@article{brito2014superradiant,
  title={Superradiant instability of black holes immersed in a magnetic field},
  author={Brito, Richard and Cardoso, Vitor and Pani, Paolo},
  journal={Physical Review D},
  volume={89},
  number={10},
  pages={104045},
  year={2014},
  publisher={APS},
  eprint={1405.2098},
  archivePrefix={arXiv},
  primaryClass={gr-qc},
  doi={10.1103/PhysRevD.89.104045}
}

@article{santos2021black,
  title={Black holes, stationary clouds and magnetic fields},
  author={Santos, Nuno M. and Herdeiro, Carlos A. R.},
  journal={Physics Letters B},
  volume={815},
  pages={136142},
  year={2021},
  publisher={Elsevier},
  eprint={2102.04989},
  archivePrefix={arXiv},
  primaryClass={gr-qc},
  doi={10.1016/j.physletb.2021.136142}
}

@article{siahaan2025melvinkerrclouds,
  title={Scalar clouds around Melvin--Kerr black holes},
  author={Siahaan, Haryanto M.},
  journal={The European Physical Journal C},
  volume={85},
  number={9},
  pages={988},
  year={2025},
  month={sep},
  publisher={Springer Nature},
  doi={10.1140/epjc/s10052-025-14719-4}
}

@article{siahaan2025melvinkerrsuperradiance,
  title={Superradiant bound states of charged scalars around Melvin--Kerr black holes},
  author={Siahaan, Haryanto M.},
  journal={Physics Letters B},
  volume={870},
  pages={139925},
  year={2025},
  doi={10.1016/j.physletb.2025.139925}
}

@article{podolsky2025kerr,
  title={Kerr black hole in a uniform {Bertotti-Robinson} magnetic field: An exact solution},
  author={Podolsk{\'y}, Ji{\v{r}}{\'\i} and Ovcharenko, Hryhorii},
  journal={Physical Review Letters},
  volume={135},
  number={18},
  pages={181401},
  year={2025},
  eprint={2507.05199},
  archivePrefix={arXiv},
  primaryClass={gr-qc},
  doi={10.1103/rfgv-ybz5}
}

@book{abramowitz1964handbook,
  title={Handbook of mathematical functions with formulas, graphs, and mathematical tables},
  author={Abramowitz, Milton and Stegun, Irene A},
  volume={55},
  year={1964},
  publisher={US Government printing office}
}

@book{bender1999advanced,
  title={Advanced mathematical methods for scientists and engineers I: Asymptotic methods and perturbation theory},
  author={Bender, Carl M and Orszag, Steven A},
  year={1999},
  publisher={Springer Science \& Business Media}
}

@book{press2007numerical,
  title={Numerical recipes 3rd edition: The art of scientific computing},
  author={Press, William H and Teukolsky, Saul A and Vetterling, William T and Flannery, Brian P},
  year={2007},
  publisher={Cambridge university press}
}

@article{bardeen1972rotating,
  title={Rotating black holes: locally nonrotating frames, energy extraction, and scalar synchrotron radiation},
  author={Bardeen, James M and Press, William H and Teukolsky, Saul A},
  journal={The Astrophysical Journal},
  volume={178},
  pages={347--370},
  year={1972},
  doi={10.1086/151796}
}

@article{zhang2024energy,
  title={Energy extraction via magnetic reconnection in magnetized black holes},
  author={Zhang, Shao-Jun},
  journal={Journal of Cosmology and Astroparticle Physics},
  volume={2024},
  number={7},
  pages={042},
  year={2024},
  month={jul},
  publisher={IOP Publishing and SISSA},
  eprint={2405.16941},
  archivePrefix={arXiv},
  primaryClass={gr-qc},
  doi={10.1088/1475-7516/2024/07/042}
}

@article{Galtsov:1978ag,
    author = "Galtsov, D. V. and Petukhov, V. I.",
    title = "{Black Hole in an External Magnetic Field}",
    journal = "Zh. Eksp. Teor. Fiz.",
    volume = "74",
    pages = "801--818",
    year = "1978"
}

@misc{Hu:2026slp,
    author = "Hu, Li and Cai, Rong-Gen and Wang, Shao-Jiang",
    title = "{Thermodynamics of Kerr-Bertotti-Robinson black hole}",
    howpublished = "arXiv:2603.18821 [gr-qc]",
    eprint = "2603.18821",
    archivePrefix = "arXiv",
    primaryClass = "gr-qc",
    year = "2026"
}

@article{PhysRev.116.1331,
  title = {Uniform Electromagnetic Field in the Theory of General Relativity},
  author = {Bertotti, B.},
  journal = {Phys. Rev.},
  volume = {116},
  issue = {5},
  pages = {1331--1333},
  numpages = {0},
  year = {1959},
  month = {Dec},
  publisher = {American Physical Society},
  doi = {10.1103/PhysRev.116.1331},
  url = {https://link.aps.org/doi/10.1103/PhysRev.116.1331}
}

@article{robinson1959solution,
  title={A solution of the Maxwell-Einstein equations},
  author={Robinson, I.},
  journal={Bulletin de l'Acad{\'e}mie Polonaise des Sciences. S{\'e}rie des Sciences Math{\'e}matiques, Astronomiques et Physiques},
  volume={7},
  pages={351--352},
  year={1959}
}

@article{Gray2026,
  author = {Gray, Finnian and Kubiz{\v{n}}{\'a}k, David and Ovcharenko, Hryhorii and Podolsk{\'y}, Ji{\v{r}}{\'\i}},
  title = {Hidden symmetries and separability structures of {Ovcharenko--Podolsk{\'y}} and conformal-to-{Carter} spacetimes},
  journal = {Phys. Rev. D},
  volume = {113},
  pages = {044050},
  year = {2026},
  eprint = {2511.21538},
  archivePrefix = {arXiv},
  primaryClass = {gr-qc},
  doi = {10.1103/PhysRevD.113.044050}
}

@article{EHT2021Magnetic,
  author = {{Event Horizon Telescope Collaboration}},
  title = {First {M87} Event Horizon Telescope Results. {VIII}. Magnetic Field Structure near the Event Horizon},
  journal = {Astrophys. J. Lett.},
  volume = {910},
  number = {1},
  pages = {L13},
  year = {2021},
  eprint = {2105.01173},
  archivePrefix = {arXiv},
  primaryClass = {astro-ph.HE},
  doi = {10.3847/2041-8213/abe4de}
}

@article{Wang2026Shadows,
  author = {Wang, Xinyu and Hou, Yehui and Wan, Xi and Guo, Minyong and Chen, Bin},
  title = {Geodesics and shadows in the {Kerr--Bertotti--Robinson} black hole spacetime},
  journal = {JCAP},
  volume = {02},
  pages = {050},
  year = {2026},
  eprint = {2507.22494},
  archivePrefix = {arXiv},
  primaryClass = {gr-qc},
  doi = {10.1088/1475-7516/2026/02/050}
}

@misc{Wang2026M87,
  author = {Wang, Chao-Hui and Meng, Xiang-Cheng and Wei, Shao-Wen},
  title = {Magnetic field effects on spherical orbit in {Kerr--Bertotti--Robinson} spacetime: Constraints from jet precession of {M87*}},
  year = {2026},
  howpublished = {arXiv:2602.03161 [gr-qc]},
  eprint = {2602.03161},
  archivePrefix = {arXiv},
  primaryClass = {gr-qc}
}

\end{document}